\documentclass[aps,prd,superscriptaddress,nofootinbib,twocolumn,notitlepage,longbibliography,10pt]{revtex4-1}

\usepackage{amsmath,amssymb,amsbsy,amstext,amsfonts,mathrsfs,latexsym,bm,physics}
\usepackage[colorlinks=true,urlcolor=blue,anchorcolor=blue,citecolor=blue,linkcolor=blue,linktocpage=true,pdfa=true]{hyperref}
\usepackage{graphicx}

\begin{document}

\title{Microphysical manifestations of viscosity and consequences for anisotropies in the very early universe}

\author{Chandrima Ganguly}
\email{c.ganguly@damtp.cam.ac.uk}
\affiliation{Department of Applied Mathematics and Theoretical Physics, Centre for Mathematical Sciences, University of
Cambridge, Wilberforce Road, Cambridge CB3 0WA, United Kingdom}
\affiliation{Wolfson College, Barton Road, Cambridge CB3 9BB, United Kingdom}

\author{Jerome Quintin}
\email{jerome.quintin@aei.mpg.de}
\affiliation{Max Planck Institute for Gravitational Physics (Albert Einstein Institute), D-14476 Potsdam, Germany}

\begin{abstract}
It has been known that a non-perfect fluid that accounts for dissipative viscous effects can evade a highly anisotropic chaotic mixmaster approach to a singularity. Viscosity is often simply parameterised in this context, so it remains unclear whether isotropisation can really occur in physically motivated contexts. We present a few examples of microphysical manifestations of viscosity in fluids that interact either gravitationally or, for a scalar field for instance, through a self-coupling term in the potential. In each case, we derive the viscosity coefficient and comment on the applicability of the approximations involved when dealing with dissipative non-perfect fluids. Upon embedding the fluids in a cosmological context, we then show the extent to which these models allow for isotropisation of the universe in the approach to a singularity. We first do this in the context of expansion anisotropy only, i.e., in the case of a Bianchi type-I universe. We then include anisotropic 3-curvature modelled by the Bianchi type-IX metric. It is found that a self-interacting scalar field at finite temperature allows for efficient isotropisation, whether in a Bianchi type-I or type-IX spacetime, although the model is not tractable all the way to a singularity. Mixmaster chaotic behaviour, which is well known to arise in anisotropic models including anisotropic 3-curvature, is found to be suppressed in the latter case as well. We find that the only model permitting an isotropic singularity is that of a dense gas of black holes.
\end{abstract}

\maketitle

\section{Introduction}

While the currently observable universe is isotropic to a very high degree, this is not a generic feature of spacetimes near singularities --- rather the opposite. In fact, within general relativity and under certain assumptions in the matter sector, the most generic approaches to singularities (such as cosmological big bang or big crunch singularities, but also black hole singularities) are highly anisotropic. They display a chaotic mixmaster behaviour \cite{Misner:1969hg}, with infinite chaotic oscillations on a finite time interval. This behaviour proceeds in epochs, the beginning and end of which are well approximated by the Kasner metric \cite{Kasner:1921zz}. This is known as the Belinski-Khalatnikov-Lifshitz (BKL) singularity \cite{Belinsky:1970ew} (see, e.g., \cite{Belinski:2017fas,Belinski:2009wj,Belinski:2014kba} for reviews).

In the context of the current paradigm of very early universe cosmology, anisotropies from the big bang initial singularity would be quickly washed away by a period of accelerated expansion (i.e.~inflation). However, our knowledge of possible pre-inflationary physics is very scarce, and the question of what happened near the big bang remains of fundamental interest. In particular, semi-classical general relativity most likely does not hold anymore at such high energy scales, and what was the nature of the initial big bang singularity (if there was one) remains an open question, especially whether it was of BKL type or rather isotropic. Within string theory, chaotic anisotropies are expected (e.g., \cite{Damour:2000wm,Damour:2000hv,Damour:2002tc,Damour:2002et}), while some semi-classical higher-derivative theories of gravity have stable isotropic cosmological singularities (e.g., \cite{Middleton:2008rh}) or can limit the growth of anisotropies \cite{Barrow:2005qv,Barrow:2006xb}. Other theories of modified gravity can similarly bound shear anisotropies \cite{Sakakihara:2020rdy} or screen them \cite{Starobinsky:2019xdp,Galeev:2021xit}. Moreover, in a gravitational ultraviolet-complete theory such as quadratic gravity \cite{Stelle:1976gc}, requiring semi-classical cosmological transition amplitudes from the big bang to today to be well defined and finite severely constrains anisotropic singularities \cite{Lehners:2019ibe,Jonas:2021xkx}.

In the context of alternative very early universe scenarios such as models of bouncing cosmology, the question of the evolution of anisotropies also plays an important role, even well before the approach to the high-curvature big crunch/bounce singularity (or before a non-singular bounce occurs). For instance, in a Bianchi type-I universe, the contribution from shear anisotropies to the total energy density budget is proportional to $1/a^6$, where $a$ is the spatially averaged scale factor. While this decays very rapidly in an expanding universe, it conversely grows much faster than for other known matter types (e.g., pressureless dust and radiation), thus representing an instability to standard isotropic contracting models. This is not an issue for ekpyrotic cosmology since the isotropic background scaling solution arises from a scalar field with energy density growing as $1/a^{2\epsilon}$ with $\epsilon>3$, thus effectively diluting anisotropies. As such, the ekpyrotic scenario has been shown to be very robust with respect to dynamically producing an isotropic universe, as demonstrated by analytic and numerical studies \cite{Erickson:2003zm,Garfinkle:2008ei,Cook:2020oaj,Ijjas:2020dws,Ijjas:2021gkf,Ijjas:2021wml}. In fact, formally, ekpyrosis implies a no-hair theorem stating that the future big crunch is a stable isotropic singularity \cite{Lidsey:2005wr}. The theorem does not hold, however, if instead of an ekpyrotic scalar field one has an imperfect fluid with anisotropic pressures that satisfy ekpyrotic equations of state \cite{Barrow:2015wfa} (deviations from perfect fluids will be further discussed below). It is also to be noted that single-field ekpyrosis predicts a blue spectrum of scalar perturbations. This is resolved in the case of two-field ekpyrosis at the cost of introducing an additional degree of freedom.

In the context of matter bounce cosmology, where a scale-invariant power spectrum of adiabatic curvature perturbations is generated during a matter-dominated contracting phase \cite{Wands:1998yp,Finelli:2001sr,Brandenberger:2012zb}, the growth of anisotropies represents a serious problem \cite{Levy:2016xcl} (to the opposite of ekpyrotic cosmology), which prevents the simplest models from being viable. Of the very few possible resolutions to this problem, we can mention the hypothetical possibility of promoting the graviton to a massive spin-2 field with mass larger than the Hubble scale in the contracting phase \cite{Lin:2017fec}. Therefore, in matter bounce cosmology and in a more general context of a bouncing or cyclic universe (not ekpyrotic), the question of how could the universe become isotropic enough for some structure formation scenario to successfully work and/or for a non-singular bounce to be achieved\footnote{Most realisations of a non-singular bounce usually simply rely on the presumption of isotropy. However, non-singular bounces with sizable anisotropies are possible (see, e.g., \cite{Bramberger:2019zez,Anabalon:2019equ,Kumar:2021mgc,Rajeev:2021yyl}), but anisotropies should be at least small enough after the bounce at the onset of radiation-dominated expansion to match later observational constraints from the cosmic microwave background \cite{Planck:2018jri}. Furthermore, if we include curvature in our bouncing model, significant anisotropies close to the bounce may not allow the universe to re-expand and create a singularity in the Weyl curvature tensor.} remains mostly unsolved.

Most approaches to cosmology from an effective field theory point of view often assume the matter content to be represented by minimally coupled scalar fields or perfect fluids. However, non-viscous fluids are an approximation to more realistic fluid dynamics models. For instance, scalar fields non-minimally coupled to gravity \cite{Faraoni:2021lfc,Giusti:2021sku}, neutrinos that are free streaming \cite{Misner:1967zz,Misner:1967uu,Stewart:1968,Matzner:1969,Weinberg:1971mx,Matzner:1972b,Weinberg:2003ur}, or any realistic interacting fluid all depict some form of viscosity. Therefore, the influence of viscosity on early- and late-time cosmologies has been studied in the context of Refs.~\cite{Hawking:1966qi,Misner:1967zz,Misner:1967uu,Stewart:1968,Stewart:1969,Matzner:1969,Weinberg:1971mx,Matzner:1972,Matzner:1972b,Parnovskii:1977,Belinskii:1979,Gron:1990ew,Weinberg:2003ur,Hervik,Brevik:review,Brevik:2019yma,Anand:2017wsj,Goswami:2016tsu,Lu:2018smr,Atreya:2017pny,Natwariya:2019fif,Mishra:2020onx} and many more. In particular, non-singular solutions have been found in the context of bulk viscosity (see, e.g., \cite{Brevik:review}), and the effect of shear viscosity has been studied as an isotropisation mechanism \cite{Belinski:2013jua,Belinski:2017fas,Ganguly:2019llh,Ganguly:2020daq}. However, the formulation of the shear viscosity term in analytic form --- while accounting for relativistic effects --- is challenging. Eckart \cite{Eckart:1940te} and Landau-Lifshitz \cite{LandauLifshitz} formulate a hydrodynamic relativistic theory of shear viscosity for models whose characteristic motion timescales are much larger than the relaxation time of the system to equilibrium. Close to a singularity, most characteristic motion should cease, so this approximation would not apply. This situation is applicable to the case of a contracting universe close to a bounce when the anisotropy energy density would grow the fastest in the absence of any other isotropising mechanism. Moreover, the theory formulated by Eckart and by Landau-Lifshitz allows for the superluminal propagation of viscous excitations. The Israel-Stewart \cite{Israel:1979wp} theory is able to rid the formalism of this problem. For the purposes of this work, we will be using a restricted version of the Israel-Stewart formalism to model the shear viscosity term. The restriction will apply in that we assume that the relaxation time to equilibrium for the fluid under consideration is very small. There exists no closed form for the viscosity term for non-zero relaxation times.

With these considerations, one can arrive at a phenomenological model for the coefficient of viscosity $\eta$ as a power law of the energy density $\rho$, i.e., $\eta \propto \rho^n$. This was the approach of previous studies, e.g., \cite{Belinski:2013jua,Belinski:2017fas,Ganguly:2019llh,Ganguly:2020daq} in the context of approaches to singularities\footnote{The literature of phenomenological studies of viscosity in \emph{general} cosmological contexts is too vast to mention here.}, but this still remains a phenomenological model. There is no microscopic model --- analogous to the kinetic theory picture of colliding hard spheres --- of the origin of this viscosity for a cosmological model. It is our intention in this work to provide the beginnings of such a microscopic realisation for viscosity embedded in concrete cosmological scenarios.

Two main avenues will be explored: an interacting scalar field in a thermal bath and black holes. The former has been extensively studied in quantum field theory (QFT), with sophisticated techniques to compute the viscosity coefficient (see, e.g., \cite{Jeon:1994if,Jeon:1995zm,Kapusta:2006pm}). Also, strongly interacting QFTs often have gravity duals (in a holographic description), from which computations have led to a viscosity bound conjecture (see, e.g., \cite{Policastro:2001yc,Kovtun:2004de,Son:2007vk}). This conjecture implies that realistic, interacting fluids always have a minimal amount of viscosity, at least of the order of their entropy density. All of this motivates us to consider a simple QFT in a cosmological background as a first microphysical realisation of viscous cosmology.

The second avenue involves black holes, which are often ubiquitous in cosmological scenarios involving a phase of contraction prior to a bounce. Indeed, black holes could form from direct collapse of inhomogeneities \cite{Banks:2002fe,Quintin:2016qro,Chen:2016kjx} or already exist from preexisting structures (as in a cyclic universe). Such black holes are expected to potentially dominate the universe near a big crunch or bounce (except possibly in regions which could undergo ekpyrotic contraction \cite{Lehners:2008qe,Lehners:2009eg}), and as such, a dense `gas' of black holes has been proposed as a state of matter at very high energies, as studied in string theory (see, e.g., \cite{Masoumi:2014vpa,Masoumi:2015sga,Masoumi:2014nfa,Mathur:2020ivc}, as well as \cite{Banks:2001px,Banks:2003ta,Banks:2004cw,Banks:2004vg} for the related holographic scenario and \cite{Veneziano:2003sz,Quintin:2018loc} for string-size black holes). In the context of black holes forming in a contracting universe, there is a serious possibility that such black holes could persist through a bounce, thus transitioning into our expanding universe as primordial black holes \cite{Carr:2011hv,Clifton:2017hvg,Carr:2017wkz,Coley:2020ykx} or remnants thereof \cite{Rovelli:2018hbk,Rovelli:2018hba,Barrau:2021spy}. The important novelty of this work is in realising that black holes, due to their gravitational attraction and intrinsic non-deformability \cite{LeTiec:2020spy,Chia:2020yla,Charalambous:2021mea}, can be treated collectively as a non-perfect fluid with shear viscosity. Therefore, under certain approximations where the hydrodynamical approximation is valid, dissipative effects form, which tend to isotropise the cosmology.

\paragraph*{Outline} We shall begin in Sec.~\ref{sec:review} by reviewing the concepts of stress, shear, viscosity, and their phenomenological implications for anisotropic cosmologies, with an emphasis on the models that are later studied in this paper. We then demonstrate in Sec.~\ref{sec:viscomani} some microphysical examples of shear viscosity: the case of an interacting scalar field theory at finite temperature and a gravitationally interacting gas of black holes, both in its dilute and dense limit. We study the effect of the viscosity coefficients derived in these scenarios on the small and the large anisotropy limits of the background universe in Sec.~\ref{sec:evo}. We briefly comment on the implications for gravitational waves in Sec.~\ref{sec:GWs}, and finally in Sec.~\ref{sec:conclusions}, we present our conclusions.

\paragraph*{Notation} Throughout this paper, we use the mostly plus metric signature $(-,+,+,+)$. Latin indices at the beginning of the alphabet run over spacetime coordinates ($a,b,c,d,\ldots\in\{0,\ldots,3\}$), while Latin indices from roughly the third of the alphabet run over spatial coordinates only ($i,j,k,\ldots\in\{1,2,3\}$). We also work with units where the speed of light, the reduced Planck constant, and the Boltzmann constant are set to unity ($c=\hbar=k_\mathrm{B}=1$), and $M_\mathrm{Pl}^2:=1/(8\pi G_\mathrm{N})$ defines the reduced Planck mass in terms of the Newtonian constant of gravitation $G_\mathrm{N}$.

\section{Review of stress, shear and viscosity}\label{sec:review}

\subsection{The definition of the shear and stress-energy tensors and the meaning of viscosity}\label{subsec:microscopic_def}

Spatially homogeneous, anisotropic models can be investigated using the orthonormal frame formalism from dynamical systems analysis (see, e.g., \cite{Ehlers:1993gf,ellis_maartens_maccallum_2012}). The geometry is split into a fluid moving orthogonally to the homogeneous spatial hypersurface, with the timelike fluid 4-velocity $u^a$ being equal to the unit normal vector of the spatial hypersurface, hence $g_{ab}u^au^b=-1$. In the spirit of the $3+1$ decomposition of the spacetime manifold, the fluid velocity vector that defines the foliation can be used to find a projection tensor,
\begin{equation}
    h_{ab}=g_{ab}+u_au_b\,,
\end{equation}
which represents the induced metric on the spatial hypersurface. The corresponding extrinsic curvature of the spatial hypersurface is then given by
\begin{equation}
    K_{ab}=h_a{}^c h_b{}^d\nabla_du_c=:\mathrm{D}_bu_a\,,
\end{equation}
where the last equality defines the spatial covariant derivative, i.e., the spacetime covariant derivative projected on the spatial hypersurface.
With simple tensorial algebra, the above can be used to show that the extrinsic curvature tensor can also be written as
\begin{equation}
    K_{ab}=\nabla_bu_a+u_bu^c\nabla_cu_a=\nabla_bu_a+u_b\dot u_a\,,
\end{equation}
where the time derivative of the fluid velocity $\dot u_a:=u^c\nabla_cu_a$ defines the acceleration of the fluid.
The extrinsic curvature tensor can be decomposed into an expansion tensor and a vorticity tensor as $K_{ab}=\Theta_{ab}+\omega_{ab}$, which are respectively symmetric ($\Theta_{ab}=K_{(ab)}$) and anti-symmetric ($\omega_{ab}=K_{[ab]}$). We assume throughout that the spacetime has no vorticity, so we set $\omega_{ab}\equiv 0$.
The expansion tensor can be further decomposed as
\begin{equation}\label{eq:defsheartensot}
    \Theta_{ab}=\frac{1}{3}\Theta h_{ab}+\sigma_{ab}\,,
\end{equation}
where $\Theta:=g^{ab}\Theta_{ab}=\mathrm{D}_au^a=\nabla_au^a$ is the trace part known as the expansion scalar, while the traceless part defines the shear tensor $\sigma_{ab}$ (so $g^{ab}\sigma_{ab}=0$). Gathering the above, the shear tensor can be written fully in terms of the fluid velocity as
\begin{equation}
    \sigma_{ab}=\mathrm{D}_{(b}u_{a)}-\frac{1}{3}h_{ab}\Theta\label{eq:sigmaabeq1}
\end{equation}
or alternatively as
$\sigma_{ab}=\nabla_{b}u_{a}+u_{b}\dot u_{a}-h_{ab}\Theta/3$
under the no-vorticity assumption. Albeit the shear tensor is traceless, a useful scalar characterisation of the shear is defined as $\sigma^2:=\sigma_{ab}\sigma^{ab}/2$, and this is used throughout this work.

The symmetric energy-momentum (or stress-energy) tensor for a generic fluid can be written as
\begin{equation}
    T_{ab}=\rho u_au_b+ph_{ab}+2q_{(a}u_{b)}+\pi_{ab}\,,
\end{equation}
where $\rho=u^au^bT_{ab}$ is the energy density, $p=h^{ab}T_{ab}/3$ is the pressure, $\pi_{ab}=h_{(a}{}^ch_{b)}{}^dT_{cd}-ph_{ab}$ is the anisotropic stress tensor (the components are also known as the anisotropic pressures), and $q_a=-h_a{}^bu^cT_{bc}$ is the heat conduction vector measured by an observer comoving with the fluid. For the purposes of this work, we ignore heat transfer ($q_a\equiv 0$).
The extra term for a non-perfect fluid represented by the anisotropic stress, which will soon be related to shear viscosity, has to satisfy the
constraints $\pi_{ab}=\pi_{ba}$, $g^{ab}\pi_{ab}=0$, and $u^a\pi_{ab}=0$,
by virtue of being the projected (symmetric) traceless part of the energy-momentum tensor.
Upon introducing a dissipative term in the energy-momentum tensor such as $\pi_{ab}$, one has to be aware that the fluid may deviate from its thermodynamic equilibrium, and the relaxation time $\tau$ to the equilibrium state (a.k.a.~the collision time or Maxwell time) may generally be non-zero. In such a case, there is no closed analytic expression for these viscous anisotropic pressures. Instead, they are defined via a differential equation \cite{Israel:1976tn,Belinski:2017fas},
\begin{equation}\label{eq:equation_shear_stress}
    \pi_{ab}+\tau h_a{}^ch_b{}^d\dot\pi_{cd}=-2\eta\sigma_{ab}\,,
\end{equation}
where $\eta$ is known as the viscosity coefficient.
Then, the entropy density, considering only shear viscous terms and setting the bulk modulus to zero, can be expanded as \cite{Belinskii:1979} (see also \cite{ellis_maartens_maccallum_2012})
\begin{equation}
    s=s_0+\frac{\pi_{ab}\pi^{ab}}{2\eta T}\tau+\mathcal{O}(\tau^2)=s_0+\frac{4\eta\sigma^2}{T}\tau+\mathcal{O}(\tau^2)\,,
\end{equation}
where $s_0$ represents the entropy density at equilibrium and $T$ is the fluid equilibrium temperature.
In order for a fluid description of the system to still be valid, we have to assume that the system is fairly close to the equilibrium state. This can thus be quantified by the following approximation,
\begin{equation}\label{eq:mfp_approx}
    \tau\ll\frac{2\eta Ts_0}{\pi_{ab}\pi^{ab}}\simeq\frac{T s_0}{4\eta\sigma^2}\,,
\end{equation}
where the approximate equality on the right-hand side precisely holds when $\tau$ is small.
For our purposes, we neglect $\tau$ in comparison to the equilibrium entropy density, and so the differential equation \eqref{eq:equation_shear_stress} collapses to the simpler expression
\begin{equation}
    \pi_{ab}=-2\eta\sigma_{ab}\,,\label{eq:defeta1}
\end{equation}
as can be seen in standard textbooks (e.g., \cite{LandauLifshitz,LandauLifshitz2}).
Since $\pi_{ab}$ is the (projected) symmetric traceless part of $T_{ab}$, it is natural for it to be proportional to the other projected symmetric traceless tensor defined above, namely the shear tensor.
This theory with $\tau \to 0$, though standard, may be plagued by a superluminal propagation of shear excitations --- the velocity of this propagation is given by \cite{Belinski:2017fas,LandauLifshitz2}
\begin{equation}
    c_\mathrm{s}\sim\sqrt{\frac{\eta}{\rho\tau}}\,.\label{eq:csshear}
\end{equation}
This issue is discussed further below.

Deviations from a perfect fluid are sometimes written as (ignoring heat transfer)
\begin{equation}
    T_{ab}=\rho u_au_b+\bar ph_{ab}+\Pi_{ab}\,,
\end{equation}
where $\bar p$ now denotes the perfect fluid pressure or average pressure. The deviation from a perfect fluid can then generally be written as a linear combination of the trace and traceless parts of the expansion tensor as
\begin{equation}
    \Pi_{ab}=-2\eta\sigma_{ab}-\zeta\Theta h_{ab}\,,
\end{equation}
which implies that the energy-momentum tensor becomes
\begin{equation}
    T_{ab}=\rho u_au_b+(\bar p-\zeta\Theta)h_{ab}-2\eta\sigma_{ab}\,,
\end{equation}
and hence the `total pressure' is $p=\bar p-\zeta\Theta$.
As shown in, e.g., Refs.~\cite{Weinberg:1971mx,Ehlers:1993gf,ellis_maartens_maccallum_2012}, the proportionality coefficients $\eta$ and $\zeta$ have the thermodynamical interpretation of shear viscosity and bulk viscosity, respectively. Bulk viscosity has the effect of modifying the pressure term. In the case of a flat universe, the bulk viscosity term, which is proportional to the volume expansion rate $\Theta$, can be expressed as a non-linear equation of state (EoS) $p=p(\rho)$, with the pressure being a quadratic function of energy density. Quadratic equations of state have been shown to admit non-singular bouncing solutions \cite{Bozza:2009jx,Ananda:2005xp,Ananda:2006gf,Ganguly:2019llh}. Given the relevance of anisotropic stress and shear anisotropies for this work, we are mostly concerned by the shear viscosity entering in \eqref{eq:defeta1}, hence we assume no bulk viscosity throughout ($\zeta\equiv 0$, so the `total pressure' and `perfect fluid pressure' have the same meaning, i.e., $\bar p=p$). The resulting energy-momentum tensor, $T_{ab}=\rho u_au_b+ph_{ab}-2\eta\sigma_{ab}$, is the same as motivated in the previous paragraph.

To gain some intuition about shear viscosity, let us consider a Minkowski background for the time being.
We can do this without loss of generality to derive the viscosity coefficient since it is an intrinsic property of the fluid (just like an EoS).
In other words, by the equivalence principle of general relativity, we are free to consider a locally Minkowski space to derive the properties of the fluid and later apply such properties to a curved spacetime.
Equations \eqref{eq:sigmaabeq1} and \eqref{eq:defeta1} for a Minkowski metric tell us that
\begin{equation}
    \pi_{ij}=-2\eta\sigma_{ij}=-2\eta\left(\partial_{(j}u_{i)}-\frac{1}{3}\delta_{ij}\partial_ku^k\right)\,,
\end{equation}
where we are specialising ourselves to the spatial components.\footnote{In fact, if we consider a frame where the fluid 4-velocity is constant, e.g., $u^a=(1,\vec{0})$, then $\sigma_{0b}=\sigma_{a0}=0$, i.e., the shear tensor is purely spatial. This is to be expected in complete generality since, as it is explicit from Eq.~\eqref{eq:sigmaabeq1}, the shear is a tensor that is fully projected onto the spatial hypersurface.}
Let us further consider a simplified setup in $(2+1)$ dimensions, where one has a fluid in between two infinite-dimensional plates. Let the fluid move in the $+x$ direction (in Cartesian coordinates), with a velocity that only depends on the $y$ direction, i.e., $u^i(x,y,z)=(u^x(y),0,0)$. This is the typical setup to derive the heuristic expression for a fluid's viscosity from kinetic theory first principles (see, e.g., \cite{ChapmanCowling,LeBellac,Burshtein}). In this setup, it is clear that the above relation between stress, viscosity and shear reduces to
\begin{equation}
    \pi^x{}_y=-\eta\partial_yu^x\,,
\end{equation}
hence the viscosity coefficient is the proportionality factor that relates the net momentum flux through a constant-$y$ surface to the velocity gradient of the fluid. In other words, viscosity is a measure of the rate of momentum diffusion in the fluid.

The mean distance between interactions among the fluid's microscopic constituents is characterised by the mean free path $\ell_\mathrm{mfp}$, and thus, the velocity difference of particles moving through a constant-$y$ surface is proportional to $-\ell_\mathrm{mfp}\partial_yu^x$. This assumes that the mean free path is much smaller than the overall size of the system, i.e., $\ell_\mathrm{mfp}\ll L$, where $L$ is the distance between the two plates in this simplified setup.
The momentum flux is then proportional to multiplying $-\ell_\mathrm{mfp}\partial_yu^x$ by the energy density $\rho$ and the mean propagation speed of the particles or the root-mean-square speed for a given statistical distribution; for the purpose of our work, as an approximation, we will simply associate this speed with the sound speed $c_\mathrm{s}$.
Combining the above, we arrive at the expression
\begin{equation}
    \eta=\alpha c_\mathrm{s}\rho\ell_\mathrm{mfp}\sim\frac{c_\mathrm{s}E}{\sigma_\mathrm{cs}}\,,\label{eq:etakin}
\end{equation}
where $\alpha$ is a proportionality constant of $\mathcal{O}(1)$ whose precise value depends on the exact microphysics at play, the statistical distribution, etc.; moreover, it shall encapsulate the uncertainty in our choice of mean velocity.
In the second equality above (up to a proportionality factor of order unity, hence the sign $\sim$), we used the fact that we can write the mean free path as $\ell_\mathrm{mfp}=1/(\beta n\sigma_\mathrm{cs})$ in terms of the number density $n$, related to the energy density by the energy of the individual particles $E$ via $\rho=En$, and the cross sectional area $\sigma_\mathrm{cs}$, which is a measure of the interaction probability among particles. The constant proportionality factor $\beta$ of order unity again depends on the exact statistical distribution.

From the above, we see that the smaller the interaction probability (i.e., the smaller the cross section), the farther a particle travels before interacting with another one (i.e., the larger the mean free path), the easier the momentum transfer, and therefore the larger the viscosity is. However, one needs to be careful since it would appear the limit $\sigma_\mathrm{cs}\to 0$ implies infinite viscosity, when one would rather believe that a fluid with no interactions should be viscous-free. Indeed, the issue with the vanishing cross section limit is that it implies an infinite mean free path, hence the assumption $\ell_\mathrm{mfp}\ll L$ is broken. In cosmology, the size of the system of interest can be associated with the Hubble radius, $L\sim|H|^{-1}$. We shall thus be particularly careful with this assumption throughout this work in order to remain in the regime of validity for the expression \eqref{eq:etakin} to hold. Nevertheless, if $\ell_\mathrm{mfp}\sim L$, it does not mean that viscosity goes away. In fact, the approximation can often be pushed to that limit within order 1 corrections that slightly reduce the viscosity (see, e.g., \cite{ChapmanCowling}). However, when $\ell_\mathrm{mfp}\gg L$, the above expression for viscosity definitely breaks down, and one generally expects $\eta\to 0$ as $\ell_\mathrm{mfp}\to\infty$.

Let us mention that the mean free path and the relaxation time (the average time between collisions) are related by the average velocity: $\ell_\mathrm{mfp}\sim c_\mathrm{s}\tau$. Hence, one can see that \eqref{eq:csshear} and \eqref{eq:etakin} are consistently related. This allows us to re-express the approximation \eqref{eq:mfp_approx} as an upper bound on the mean free path,
\begin{equation}
    \ell_\mathrm{mfp}\lesssim\sqrt{\frac{Ts_0}{\rho\sigma^2}}=:\ell_\mathrm{max}\,.\label{eq:deflmax}
\end{equation}
Moreover, one could demand the speed of propagation not to surpass the speed of light, which from \eqref{eq:csshear} amounts to a lower bound on the mean free path. Combining those, and from the discussion of the previous paragraph, we arrive at the following regime of validity:
\begin{equation}
    \frac{\eta}{\rho}\lesssim\ell_\mathrm{mfp}\lesssim\mathrm{min}\left\{\ell_\mathrm{max},|H|^{-1}\right\}\,.\label{eq:validitymfp}
\end{equation}
Therefore, given a model for which one can compute the viscosity thanks to Eq.~\eqref{eq:etakin}, the above lower and upper bounds essentially tell us the regime of validity of that expression in terms of the size of the fluid's mean free path. This shall be the basis of our consistency checks throughout this work.

\subsection{The effect of shear viscosity in anisotropic cosmology}\label{sec:viscoanicosmo}

In order to study the effect of shear viscosity of the form \eqref{eq:defeta1} in cosmology, let us write down
the Einstein equations with no cosmological constant, $G_{ab}=M_\mathrm{Pl}^{-2}T_{ab}$, as follows when $q_a=\omega_{ab}=\dot u_a=0$ \cite{ellis_maartens_maccallum_2012},
\begin{subequations}\label{eq:EFE-orthonormal}
\begin{align}
    \frac{1}{3}\Theta^2&=\frac{\rho}{M_\mathrm{Pl}^2}-\frac{1}{2}{}^{(3)}\!R+\sigma^2\,,\\
    \dot\Theta+\frac{1}{3}\Theta^2&=-\frac{1}{2M_\mathrm{Pl}^2}(\rho+3p)-2\sigma^2\,,\\
    \dot\rho+\Theta(\rho+p)&=-\pi^{ab}\sigma_{ab}\,,\\
    \dot\sigma_{ab}+\Theta\sigma_{ab}&=\frac{1}{M_\mathrm{Pl}^2}\pi_{ab}-{}^{(3)}\!R_{ab}+\frac{1}{3}{}^{(3)}\!Rh_{ab}\,,
\end{align}
\end{subequations}
where ${}^{(3)}\!R_{ab}$ and ${}^{(3)}\!R$ are, respectively, the 3-dimensional Ricci curvature tensor and scalar on the spatial hypersurface. This system has within it cosmologies containing anisotropies in the expansion, i.e.~different expansion rates in the $3$ different spatial directions, as well as containing anisotropies in the $3$-curvature.

In order to study the effects of anisotropic pressure on an anisotropic universe, let us specialise to a simple flat anisotropic universe. This is known as the Bianchi type-I universe. It represents the case of maximal anisotropy when it is empty, in which case it is called the Kasner solution. It also only has expansion anisotropy, instead of anisotropy in the $3$-curvature as well. The metric can be represented as
\begin{equation}\label{eq:BIMetric}
    g_{ab}\dd x^a\dd x^b=-\dd t^2+a(t)^2e^{2\beta_{(i)}(t)}\delta_{ij}\dd x^i\dd x^j\,,
\end{equation}
with the constraint $\sum_{i=1}^3\beta_{(i)}(t)=0$, where $\beta_{(i)}(t)$ denotes the anisotropy in direction $x^i$ and $a(t)$ denotes the spatially averaged scale factor (in the sense that $\ln a=\langle\ln a_{(i)}\rangle$ with $a_{(i)}=ae^{\beta_{(i)}}$). From this, $H(t):=\dot a/a$ defines the spatially averaged Hubble parameter, and the hypersurface geometry is characterized by ${}^{(3)}\!R_{ab}={}^{(3)}\!R=0$, $\Theta=3H$, and
\begin{equation}
    \sigma_{ij}=a^2e^{2\beta_{(i)}}\dot\beta_{(i)}\delta_{ij}\,,\qquad\sigma_i{}^j=\dot\beta_{(i)}\delta_i{}^j\,.
\end{equation}
In particular, $\sigma^2=(1/2)\sum_{i=1}^3\dot\beta_{(i)}^2$. The resulting equations of motion (EOMs) are
\begin{subequations}\label{eq:BIall}
\begin{align}
    3M_\mathrm{Pl}^2H^2&=\rho+\rho_\sigma\,,\label{eq:BIconstrgen}\\
    2M_\mathrm{Pl}^2\dot H&=-(\rho+p)-2\rho_\sigma\,,\label{eq:BIHEOM}\\
    \dot\rho+3H(\rho+p)&=-\pi^{ab}\sigma_{ab}\,,\label{eq:BIrhoEOM}\\
    \dot\sigma_{ab}+3H\sigma_{ab}&=M_\mathrm{Pl}^{-2}\pi_{ab}\,,\label{eq:BIsigmaEOM}
\end{align}
\end{subequations}
where $\rho_\sigma:=M_\mathrm{Pl}^2\sigma^2$ defines the energy density in shear anisotropies.

In the presence of a perfect fluid, the stress tensor vanishes, and we recover the shear equation\footnote{Note that, while $\dot f=\partial_tf$ for any scalar-valued function $f$, we have $\dot\sigma_{ij}=u^a\nabla_a\sigma_{ij}=\partial_t\sigma_{ij}-2(H+\dot\beta_{(i)})\sigma_{ij}$ for a rank-2 tensor in the above Bianchi type-I spacetime. Also, recall $\sigma_{ab}$ is purely spatial, so in particular $\sigma^2=\sigma_{ab}\sigma^{ab}/2=\sigma_{ij}\sigma^{ij}/2$.}
\begin{equation}
    \partial_t\sigma_i{}^j+3H\sigma_i{}^j=0\implies\sigma_i{}^j\propto\frac{1}{a^3}\,,
\end{equation}
and so $\rho_\sigma\propto 1/a^6$, according to which shear anisotropies essentially contribute to the Friedmann equations as a perfect fluid with stiff EoS $p_\sigma=\rho_\sigma$. In particular, one recovers the result that anisotropies typically dominate the energy budget of the universe near cosmological singularities since, as $a\to 0$, $\rho_\sigma\propto 1/a^6$ grows faster than $\rho\propto 1/a^{-3(1+w)}$ for a background perfect fluid with EoS $w:=p/\rho\in(-1,1)$. As a result, the spacetime near the singularity is well approximated by the anisotropic Kasner metric, and the approach to the metric is of BKL type, as mentioned in the Introduction.

An immediate loophole is if the matter EoS satisfies $w>1$, which is known as an ultra-stiff ekpyrotic EoS, in which case the background energy density dominates as the scale factor goes to small values, hence isotropising the universe such that it becomes well approximated by a Friedmann-Lema\^itre-Robertson-Walker (FLRW) metric.
In the same situation, the general existence of anisotropic stresses acts as a positive source, and the shear equation of motion is modified according to Eq.~\eqref{eq:BIsigmaEOM}.
This causes the energy density in the anisotropies to grow faster than $a^{-6}$, and hence an ekpyrotic fluid can no longer be reliably expected to isotropise the universe. On doing an extension of this study to anisotropic cosmologies with anisotropic $3$-curvature as well as expansion anisotropies (for example, in Bianchi type IX), one finds that anisotropic stresses even if they are ultra-stiff on average, fail to isotropise the cosmology. In fact, a bounce fails to occur as the geometry approaches an anisotropic singularity \cite{Barrow:2015wfa}.

Let us now try to gain some intuition about how shear viscosity might change this picture. This was discussed initially in the context of neutrino viscosity and its effects on isotropisation \cite{Misner:1967uu}. The discussion was extended to derive a possible phenomenological form of such a shear viscous term in \cite{Belinski:2017fas}. There, one postulates a shear viscosity coefficient in an anisotropic stress of the form \eqref{eq:defeta1}, which is dependent on a power law of the energy density as $\eta\propto\rho^{1/2}$. The power is $1/2$, which allows for isotropisation and an attractor behaviour to a Friedmann singularity \cite{Belinski:2017fas}. This form of the viscous anisotropic stress, though, allows for the propagation of super-luminal excitations, which we have at the cost of the viscous stresses having a closed form.

If the shear viscosity enters the stress tensor as in Eq.~\eqref{eq:defeta1}, then the matter conservation equation and the shear EOM are generally modified as follows:
\begin{subequations}\label{eq:mattershearEOMs}
\begin{align}
    \dot\rho+3H(\rho+p)&=4\eta\sigma^2\,;\\
    \partial_t\sigma_i{}^j+3H\sigma_i{}^j&=-2M_\mathrm{Pl}^{-2}\eta\sigma_i{}^j\,.\label{eq:sigmaijBI}
\end{align}
\end{subequations}
Together with Eqs.~\eqref{eq:BIconstrgen}--\eqref{eq:BIHEOM}, those are typically coupled, first-order ordinary differential equations (not necessarily linear), for which analytic solutions can be found only in special cases.
Moreover, the viscosity coefficient is in general time dependent (i.e., it may depend on background quantities such as $a$, $\rho$, $H$, etc.).

Let us first consider the simplest case where it is simply a constant, i.e., $\eta=\mathrm{constant}=:\kappa$, with mass dimension $3$. We use a different variable $\kappa$ here to denote the constant viscosity since we will use such a positive, dimensionful\footnote{The dimensionality of $\kappa$ depends on the expression; it may not always be the same.} constant of proportionality for the viscosity coefficient throughout, i.e., it will serve as a reference scale in the time-dependent examples below.
Accordingly, the shear EOM becomes
\begin{equation}
    \partial_t\sigma_i{}^j+3\left(\frac{\dot a}{a}\right)\sigma_i{}^j+2\left(\frac{\kappa}{M_\mathrm{Pl}^2}\right)\sigma_i{}^j=0\,,
\end{equation}
whose general solution can be written in the form
\begin{equation}
    \rho_\sigma\propto\frac{1}{a^6}\exp\left(-\frac{4\kappa t}{M_\mathrm{Pl}^2}\right)\,,\label{eq:rhosigmaconstanteta}
\end{equation}
as was already found by Misner \cite{Misner:1967zz}.
We notice that the $1/a^6$ behaviour is modified due to the constant viscosity coefficient $\kappa$ by an exponential factor in time.
Applying this to our physical considerations of interest, we note that the BKL approach to a singularity\footnote{Note that the BKL singularity is related to a singularity in the Weyl tensor which is directly related to $\sigma_{ij}$.} is probably not affected too much since, as $a\to 0$ and $t\to 0$, one gets very close to the situation where $\rho_\sigma\sim 1/a^6\to\infty$. Nevertheless, the exact growth rate of the shear anisotropies is modified in the approach to a singularity, but its exact value can only be recovered provided a solution for $a(t)$ is also found, which requires additional input.
While a constant viscosity coefficient may not isotropise a singularity, it might still dilute anisotropies over an intermediate timescale thanks to the above exponential suppression in $\rho_\sigma$. Such an example will be explored in greater detail in the subsequent section.

As a second example, let us explore the possibility that $\eta=\kappa/a^3$, which we will motivate in the next section. In fact, this will appear as a possible scaling of the viscosity coefficient in the context of an interacting scalar field theory in a radiation bath. In such a context, the shear EOM can be rewritten as
\begin{equation}
    H\left(a(\sigma_i{}^{j})'+3\sigma_i{}^j\right)+2\left(\frac{\kappa}{M_\mathrm{Pl}^2a^3}\right)\sigma_i{}^j=0\,,
\end{equation}
where a prime here denotes a derivative with respect to $a$. Assuming the background to be radiation dominated, one has $a(t)=\sqrt{t/t_0}$, and so $H(a)=1/(2t_0a^2)$ is positive for $t_0>0$ (expansion) and negative for $t_0<0$ (contraction). As a result, one can solve the above differential equation, and the evolution of shear anisotropies is modified as
\begin{equation}
    \rho_\sigma\propto\frac{1}{a^6}\exp\left(\frac{8\kappa t_0}{M_\mathrm{Pl}^2a}\right)\,.\label{eq:shearsoletaam3}
\end{equation}
Interestingly, if $t_0<0$ (contraction), one finds that $\rho_\sigma\to 0$ as $a\to 0$, and so it appears that anisotropies have been fully washed out by the time of a big crunch. Alternatively, if $t_0>0$ (expansion), one finds that $\rho_\sigma$ badly blows up in the backward approach to the big bang, exponentially more severely than in the BKL case. Equivalently, it means the anisotropies very quickly decay under forward time evolution in an expanding universe. However, in both instances (contraction and expansion), the meaning of viscosity near the singularity might be lost, as will be discussed in greater detail in the next section.

As a last example for this section, let us consider the possibility that $\eta=\kappa|H|$ (in this case, $\kappa$ has mass dimension $2$), which will be further motivated in the next section.\footnote{We note that this corresponds to the case $\eta\sim\sqrt{\rho}$ in a regime where $\rho\gg\rho_\sigma$ according to Eq.~\eqref{eq:BIconstrgen}. This is the scaling that was noticed to lead to perfect isotropisation \cite{Belinski:2017fas,Ganguly:2020daq}.} For simplicity, let us rewrite this expression as $\eta=\varepsilon\kappa H$ with $\varepsilon=+1$ for $H>0$ (expansion) and $\varepsilon=-1$ for $H<0$ (contraction). The shear EOM in this case reduces to
\begin{equation}
    \partial_t\sigma_i{}^j+\left(3+2\varepsilon\frac{\kappa}{M_\mathrm{Pl}^2}\right)H\sigma_i{}^j=0\,,
\end{equation}
whose solution is immediately read off to be
\begin{equation}
    \rho_\sigma\propto\frac{1}{a^{6+4\varepsilon\kappa/M_\mathrm{Pl}^2}}\,.\label{eq:rhosigmaDBHG}
\end{equation}
Interestingly, the growth rate of the shear anisotropies is modified in such a case by adding a correction to the power of the $1/a^6$ scaling; in fact, one can write it as $\rho_\sigma\propto 1/a^{6+\delta}$ with $\delta=4\varepsilon\kappa/M_\mathrm{Pl}^2$. One then notices that for $\varepsilon=-1$ (contraction), the shear anisotropies grow less fast than the typical behaviour in the approach to a big crunch\footnote{In fact, one even finds that $\rho_\sigma\to 0$ as $a\to 0$ if $\kappa>3M_\mathrm{Pl}^2/2$, meaning that anisotropies would be completely damped out by the time of the crunch in such a case.}, while for $\varepsilon=+1$ (expansion), they grow faster in the (backward) approach to the big bang; they also correspondingly decay faster under forward time evolution out of the big bang.

This analysis can be extended to cases where there is both expansion and curvature anisotropy. One such example is the Bianchi type-IX universe. This is the case that is taken to be the generic approach to the singularity according to the BKL analysis \cite{Belinsky:1970ew}. Due to the presence of the anisotropic $3$-curvature, this cosmology on contraction shows infinite chaotic mixmaster oscillations on a finite time interval (when the lower limit of that time interval is $0$), which is an attractor behaviour. An isotropisation mechanism that is successful would be able to resolve this attractor behaviour in the form of chaotic oscillations. In the absence of anisotropic pressures, numerical studies by \cite{Garfinkle:2008ei} among others show that ekpyrosis is successful in doing this. In a separate work \cite{Ganguly:2019llh}, a viscosity coefficient of the form $\eta=\kappa\rho^{1/2}$ for some constant $\kappa$ of mass dimension $1$ is shown to successfully isotropise a Bianchi-IX universe as well as mitigate the mixmaster chaos.

\section{Some examples of microphysical manifestations of viscosity}\label{sec:viscomani}

\subsection{Interacting scalar field theory at finite temperature}\label{subsec:finite-temp-intro}

Let us consider a finite-temperature scalar field theory with action of the form
\begin{equation}
    \label{eq:scalar}
    S=\int\dd^4x\,\sqrt{-g}\left(\frac{M_\mathrm{Pl}^2}{2}R-\frac{1}{2}g^{ab}\nabla_a\phi\nabla_b\phi-V(\phi)\right)\,,
\end{equation}
with potential
\begin{equation}\label{eq:potential_finiteTemp}
    V(\phi)=\frac{1}{2}m^2\phi^2+\frac{\lambda}{4!}\phi^4\,,
\end{equation}
where $m>0$ and $0<\lambda\ll 1$ are the mass and the self-interaction coupling constant, respectively.
The physical mass of the field is well approximated by $m(1+\mathcal{O}(\lambda))\simeq m$ at weak coupling (after renormalization, up to radiative corrections and at zero temperature).
Denoting $\mu:=m/M_\mathrm{Pl}$, one could imagine having the following hierarchy of scales: $0<\mu\ll\mu/\lambda\ll\lambda\ll 1$. This allows one to distinguish various regimes where the system behaves very differently as a function of the temperature $T$ of the thermal bath \cite{Jeon:1995zm}.
Let us emphasize two such regimes:
\begin{itemize}
    \item When $0<T/M_\mathrm{Pl}\ll\mu$, the system is effectively composed of a non-relativistic, dust-like scalar field. Indeed, the potential is dominated by the zero-temperature mass, i.e., $V(\phi)\simeq m^2\phi^2/2$. In an FLRW background, as long as $|H|\ll m$ (or $|H|/M_\mathrm{Pl}\ll\mu$ in dimensionless units), the field is coherently oscillating with vanishing time-averaged effective pressure, i.e., the EoS is that of dust. If one explores the limit of the universe getting smaller, the Hubble scale and the temperature of the thermal bath both grow as $|H|\sim\rho^{1/2}\sim a^{-3/2}$ and $T\propto a^{-1}$, so radiation with $\rho\propto a^{-4}$ will quickly become dominant. Nevertheless, the mass term in the Lagrangian remains important in intermediate regimes within $\mu\lesssim T/M_\mathrm{Pl}\lesssim\mu/\lambda$.
    \item When $T/M_\mathrm{Pl}\gg\mu/\lambda$, the mass term becomes negligible, so the potential is dominated by the interaction term, i.e.~the $\lambda\phi^4$ term. In this regime, the EoS is that of radiation, $p=\rho/3$, with energy density growing as $\rho\propto T^4\propto a^{-4}$. What is crucial is that in this regime the interactions imply a scattering cross section already at the level of the $2\to 2$ tree diagram, and consequently, the fluid should have shear viscosity. We will expand on this below.
\end{itemize}
In an anisotropic background, the anisotropies would quickly begin to dominate in the approach to a singularity for such a scalar field model, ignoring viscous effects. This is most simply seen in the case of Bianchi I, which contains only expansion anisotropies and in which the energy density in the anisotropies grows as $a^{-6}$.
However, the thermal bath would remain, and thanks to the rising temperature, the regime $T/M_\mathrm{Pl}\gg\mu/\lambda$ would be reached. From then on, the presence of the interaction term implies the appearance of viscosity, which may alter the evolution of anisotropies even if those are \textit{a priori} dominant and large. How efficient viscosity may be at damping the anisotropies will be addressed later.

Let us now consider the scalar field above dominated by its self-interaction potential of the form $\lambda\phi^4$ in a thermal bath with temperature $T$. (This follows the example discussed in Ref.~\cite{Son:2007vk}).
Then, this scalar field in a thermal bath behaves like radiation with background energy density, number density, and temperature scaling as $\rho\propto a^{-4}$, $n\propto a^{-3}$, and $T\propto a^{-1}$, respectively; in particular, $n\propto T^3$. Additionally, QFT at finite temperature gives a cross section\footnote{Some intuition for this goes as follows \cite{Jeon:1994if}: the typical cross section of a $\lambda\phi^4$ theory goes as $\sigma_\mathrm{cs}\sim\lambda^2/s$, where $s$ is the square of the center-of-mass energy here. In the limit of interest, in particular for $T\gg m$, one can argue that the only relevant energy scale is the temperature, hence $s\sim T^2$ and $\sigma_\mathrm{cs}\sim\lambda^2/T^2$.} $\sigma_\mathrm{cs}\sim\lambda^2/T^2$. Putting everything together, the mean free path is
\begin{equation}\label{eq:mfp_def}
    \ell_\mathrm{mfp}\sim\frac{1}{\lambda^2T}\,.
\end{equation}
As viscosity must be measured on scales much larger than the mean free path, it follows that one cannot take the decoupling limit, $\lambda\rightarrow 0$, at which the mean free path goes to infinity.
We recall that one should demand $\ell_\mathrm{mfp}\lesssim|H|^{-1}$ on cosmological scales. If we are in a radiation-dominated background, we have $M_\mathrm{Pl}^2H^2\sim\rho\sim a^{-4}\sim T^4$, and this would mean $T/M_\mathrm{Pl}\lesssim\lambda^2$. We would thus have to be in the regime $\mu/\lambda\ll T/M_\mathrm{Pl}\lesssim\lambda^2\ll\lambda\ll 1$, so one cannot take $\lambda$ too small for the regime to exist in the first place.
Of course, once viscosity is taken into account in the Einstein equations, the background evolution is expected to be modified, and one has to find the proper regime of validity then.

Another aspect that must be considered for the above to hold is thermalization. Indeed, the finite-temperature interaction cross section only holds if the scalar field is in thermal equilibrium with the thermal bath, which is the case as long as the interaction rate $\Gamma=n\sigma_\mathrm{cs}\langle v\rangle$ is greater than the Hubble rate $|H|$. In the high-temperature relativistic limit discussed above, the field is relativistic with unit average velocity $\langle v\rangle$, hence the interaction rate is simply equal to the inverse mean free path, $\Gamma=1/\ell_\mathrm{mfp}$. Correspondingly, the requirement for thermal equilibrium is the same as the one for the kinetic theory viscosity derivation to hold, i.e., $\ell_\mathrm{mfp}<|H|^{-1}$, which we discussed above and which we will check explicitly when solving the full set of equations in the next section. Certainly, since $\Gamma\sim T$ and $|H|\sim T^2$ in a radiation-dominated universe, one does not expect thermal equilibrium to hold up to arbitrarily high energy scales.\footnote{Other effects, however, might come into play and improve thermalization. For example, \cite{1972JETP...34.1159Z} shows that thermalization is stable to particle production as long as the universe isotropises before the minima of contraction.}
A caveat to keep in mind, however, is that this only applies as a toy model, where the scalar field $\phi$ does not couple to any other fields. In a more realistic context, the physics becomes more complicated regarding thermalization. Indeed, a gauge singlet scalar field can couple to other degrees of freedom, such as the standard-model Higgs, fermions, etc. If so, as $\phi$ acquires a large vacuum expectation value (VEV), the standard-model fields would obtain large, VEV-dependent masses, which would suppress the amplitude of the scattering processes, hence delaying chemical equilibrium and thermalization. Such discussions can be found in supersymmetry (e.g., \cite{Allahverdi:2005mz} and references therein). In our context, this implies that further analysis is certainly needed.

Since the scalar field in \eqref{eq:scalar} has a canonical kinetic term, its sound speed is unity, and correspondingly, the viscosity coefficient can be evaluated according to \eqref{eq:etakin} as
\begin{equation}
    \eta\sim\frac{T^3}{\lambda^2}\,.\label{eq:etafiniteT}
\end{equation}
The exact coefficient of proportionality is difficult to estimate, but to leading order in small $\lambda$ and small $m/T$, it is expected to be in the $\mathcal{O}(1-10^3)$ regime (see, e.g., \cite{Jeon:1994if,Jeon:1995zm,Kapusta:2006pm}).
With this expression in hand, one can then solve for the Einstein equations to determine how the shear viscosity arising from the self-interacting scalar field affects the evolution of the shear anisotropies. Equation \eqref{eq:etafiniteT} suggests $\eta\propto a^{-3}$, which as we saw in Sec.~\ref{sec:viscoanicosmo}, yields the solution \eqref{eq:shearsoletaam3} upon assuming a radiation-dominated background, according to which the universe isotropises to the future. In the next section, we will solve the equations in more generality by means of numerical methods. This will also allow us to comment more specifically on the regime of validity \eqref{eq:validitymfp} over which Eq.~\eqref{eq:etafiniteT} applies.

Since the entropy density goes as $s\propto a^{-3}\propto T^3$ for a radiation bath, we notice that \eqref{eq:etafiniteT} implies
\begin{equation}
    \frac{\eta}{s}\sim\frac{1}{\lambda^2}=\mathrm{constant}\,.
\end{equation}
For $\lambda$ at least $\lesssim 1$, this means there is a constant lower bound on the ratio of the viscosity coefficient over the entropy density, $\eta/s\gtrsim 1$. This is reminiscent of the viscosity bound conjecture (e.g., \cite{Policastro:2001yc,Kovtun:2004de,Son:2007vk}), claiming $\eta/s\geq 1/(4\pi)$, which comes from anti-de Sitter black hole solutions in various gravitational theories that are holographically dual to strongly interacting QFTs (non-perfect fluids with shear viscosity). It is thus an interesting observation that many theories suggest a lower bound on the viscosity coefficient, further motivating the investigation of this work.

\subsection{Dilute gas of black holes}

In a contracting universe that is not perfectly homogeneous, perturbative inhomogeneities grow under contraction in a similar manner to anisotropies. This growth of inhomogeneities could lead us to an endpoint where a pre-bounce early universe could be populated by a gas of black holes. It was shown in \cite{Quintin:2016qro,Chen:2016kjx} that a perfect fluid with quantum vacuum initial conditions in the asymptotic past or thermal initial conditions at a finite time inevitably end up collapsing into Hubble-size black holes at a scale that is determined by the smallness of the fluid's sound speed. If structures already exist when the universe starts contracting (such as in a cyclic context), smaller black holes form first. Initially, these black holes are dilute --- this has been modelled as each black hole being situated on a lattice in \cite{Clifton:2017hvg,Coley:2020ykx}. As the universe contracts, these black holes become denser with a contracting Hubble radius. There is then a dense limit where the Schwarzschild radius $R$ is of the size of the Hubble radius, $R\sim|H|^{-1}$. This case will be dealt with in the next subsection.

In this current subsection, we shall be interested in the possible dilute case where $R\ll |H|^{-1}$, far in the contracting phase when the universe is still very large, i.e., very far away from the putative ultimate crunching singularity. We shall model a dilute gas of black holes as a set of hard balls of radius $R$ that nevertheless attract one another gravitationally.
To that level of approximation, these could in fact just be any astrophysical objects (small elliptical galaxies, stars, etc.), which might as well populate the universe in this scenario.
Interactions
between these hard spheres would give rise to a viscous drag, very similar to that derived in kinetic theory.
Although the black holes in the dilute gas would have to coalesce to form a gas of larger black holes to ultimately enter into the dense limit $R \sim |H|^{-1}$, effects of black holes (or other astrophysical objects) coalescing is not taken into account in this analysis in the dilute limit.
In fact, we do not know exactly when the perturbations become too large as to not trust the approximations, hence we must add a word of caution. While the approximations might hold initially, it is unclear how long they may last, and this has to be taken into account when drawing conclusions.

We start by saying that the cross section for two black holes as described above to interact is given by (see, e.g., \cite{Loeb:2020lwa})
\begin{equation}
    \sigma_\mathrm{cs}\sim\left(\frac{R}{c_\mathrm{s}^2}\right)^2\,,
\end{equation}
where the sound speed $c_\mathrm{s}$ of the gas represents the average velocity of the distribution\footnote{One would generally expect a distribution of masses/radii and velocities for the gas of black holes. Here we are thus referring to $R$ and $c_\mathrm{s}$ as the mean radius and velocity, respectively. We are not making any assumption about the distribution since too many factors come into play in the formation of such a gas of black holes. Beyond idealised analytical estimates as in \cite{Quintin:2016qro,Chen:2016kjx}, this would potentially require numerical simulations, which would nevertheless be very dependent on the chosen initial conditions. Therefore, we remain agnostic about exact values for $R$ and $c_\mathrm{s}$ and treat them as free parameters.} of black holes.
We note that in the relativistic limit where $c_\mathrm{s}\to 1$ the expression for the cross section reduces to that for non-interacting hard spheres, $\sigma_\mathrm{cs}\sim R^2$, as expected. Alternatively, in the pressureless limit $c_\mathrm{s}\to 0$, the cross section tends to infinity. This is understood from the fact that if all the black holes in the gas were perfectly static (say at some initial time), they would inevitably merge in some finite time due to the infinite-range gravitational attraction between them, hence the certain collision probability.

Making use of the relation between the Schwarzschild radius and mass (upon specialising our attention to black holes), $R\sim M/M_\mathrm{Pl}^2$, the number density of the gas is related to its energy density via $n\sim\rho/(RM_\mathrm{Pl}^2)$, from which we can read the mean free path $\ell_\mathrm{mfp}\sim(n\sigma_\mathrm{cs})^{-1}$ as
\begin{equation}
    \ell_\mathrm{mfp}\sim\frac{c_\mathrm{s}^4M_\mathrm{Pl}^2}{\rho R}\,.\label{eq:mfpdiluteBHG}
\end{equation}
The viscosity can then be evaluated as
\begin{equation}
    \eta\sim\frac{c_\mathrm{s}^5M_\mathrm{Pl}^2}{R}\,,\label{eq:etafindBHG}
\end{equation}
which is just a constant since in the limit $R\ll|H|^{-1}$ one does not expect the Schwarzschild radius to be affected much by the cosmological background under the present approximations.

Let us comment on the regime of validity of the above expression for viscosity, recalling \eqref{eq:validitymfp}. The inequality $\eta/\rho\lesssim\ell_\mathrm{mfp}$, which followed from demanding a sub-luminal propagation speed of shear viscosity excitations, is satisfied provided $c_\mathrm{s}\lesssim 1$. This is not a surprise as we expect the sound speed of the dilute gas of black holes to precisely be the propagation speed of viscosity excitations, and this sound speed is certainly expected to be subluminal.

The inequality on the right-hand side of \eqref{eq:validitymfp} is less trivial though. To tackle it, let us first make the observation that for a gas of black holes to first form one generally has to be in a background that is relatively close to isotropy. This can certainly be envisioned in the context of a cyclic universe, where a prior expanding phase can efficiently isotropise the universe.
Thus, we can assume here that shear is initially subdominant, or at most of the order of the fluid's energy density, i.e., $\sigma^2\lesssim\rho/M_\mathrm{Pl}^2\sim H^2$. How the shear subsequently evolves, given the viscosity \eqref{eq:etafindBHG}, will be addressed in the following section, but we already saw that a constant viscosity coefficient can lead to temporary exponential suppression of the shear [recall \eqref{eq:rhosigmaconstanteta}]. Under the assumption that shear is subdominant, the mean free path \eqref{eq:mfpdiluteBHG} can be written as $\ell_\mathrm{mfp}\sim c_\mathrm{s}^4/(RH^2)$. The requirement that this is smaller than the Hubble radius thus reads
\begin{equation}
    c_\mathrm{s}^4\lesssim R|H|\,,\label{eq:smallcsconstr}
\end{equation}
where the right-hand side is expected to be much smaller than unity since we are considering $R\ll|H|^{-1}$. Therefore, Eq.~\eqref{eq:etafindBHG} for viscosity is expected to apply only if the sound speed is very small. While $c_\mathrm{s}$ remains at the level of a free parameter given the uncertainties stipulated earlier, one certainly does not expect black holes to have large peculiar velocities upon formation from gravitational collapse, so the above inequality does not appear unreasonable.

The other requirement from the right-hand side inequality of \eqref{eq:validitymfp}, which comes from the small Maxwell time assumption, is generally found to be less stringent than \eqref{eq:smallcsconstr} as long as the shear remains subdominant. To see this, let us express the temperature of the dilute black hole gas assuming a Maxwell-Boltzmann distribution of velocities, such that $T\sim c_\mathrm{s}^2 M$. The entropy density can also be read from the sum of the black hole's individual entropies, $s\sim n(RM_\mathrm{Pl})^2\sim\rho R$, which dominates over the `ideal gas' entropy in this context. Putting those together, we arrive at $\sqrt{Ts/(\rho\sigma^2)}\sim c_\mathrm{s}RM_\mathrm{Pl}/\sigma$, and thus the mean free path \eqref{eq:mfpdiluteBHG} is smaller than $\ell_\mathrm{max}$ as long as $c_\mathrm{s}^3\lesssim R^2\rho/(M_\mathrm{Pl}\sigma)$. If $\sigma^2\ll\rho$, this is not a very severe constraint. Even if the shear is of the order of the energy density, then the constraint reduces to $c_\mathrm{s}^3\lesssim R^2|H|M_\mathrm{Pl}$, which is generally no more restrictive than \eqref{eq:smallcsconstr} since we expect to be in a deeply sub-Planckian cosmological regime ($|H|\ll M_\mathrm{Pl}$).

\subsection{Dense gas of black holes}

A contracting universe that isotropically evolves with a dilute gas of black holes will arrive at a phase where the separation distance is of the order of their Schwarzschild radius.
As the black holes are pushed closer together, we arrive at the dense black hole gas picture (e.g., \cite{Banks:2002fe}). In this picture, the EoS resembles a stiff fluid $p=\rho$, a result that is derived from thermodynamic considerations in this current section.
In the dense black hole gas picture, every Hubble patch can be thought to be filled by a Hubble-size black hole, i.e., $R=|H|^{-1}$. As the universe keeps contracting, one expects some form of quantum instability that allows black holes to `continuously' bifurcate into smaller black holes such that the relation $R=|H|^{-1}$ holds as a function of time (this is forbidden classically \cite{Hawking:1973uf}). Though such a phase is highly hypothetical, it is not violating the second law of thermodynamics as we will see below, and it might well occur if black holes at high densities are to be replaced by stringy counterparts (see, e.g., \cite{Veneziano:2003sz,Quintin:2018loc,Masoumi:2014vpa,Masoumi:2015sga,Masoumi:2014nfa,Mathur:2020ivc}).
At the level of semi-classical gravity, such a phase would inevitably still ultimately lead to a collapse of the whole universe into a singularity, but again, this is poorly studied and new physics might well come into play.

Let us consider a region of physical volume $V$ containing $N$ black holes of Schwarzschild radius $R$, so $N\sim V/R^3$.
The total energy in the volume is then given by
$E\sim NM\sim VM_\mathrm{Pl}^2/R^2$, where $M\sim M_\mathrm{Pl}^2R$ is the Schwarzschild mass of the black holes.
One must keep in mind the following: we assume that we can describe the black holes by their usual Schwarzschild mass and radius coming from the Schwarzschild metric of a single black hole embedded in Minkowski space (i.e., asymptotically flat). This might not hold in a universe that has a possibly infinite number of black holes and that could be dynamical, but we have no good prescription in that situation, so we will stick with the usual Schwarzschild description --- more comments are to be given in the discussion section.
Then, if the entropy of each black hole is given by the Bekenstein-Hawking entropy,
the total entropy in the volume is given by $S\sim NM_\mathrm{Pl}^2R^2\sim VM_\mathrm{Pl}^2/R$.
These relations can be combined to yield $S\sim M_\mathrm{Pl}\sqrt{EV}$, or in terms of densities,
\begin{equation}
    s\sim M_\mathrm{Pl}\sqrt{\rho}\,.
\end{equation}
Using standard thermodynamic relations such as $1/T=\partial_ES$ and $p=T\partial_VS$, one finds a temperature $T\sim\sqrt{\rho}/M_\mathrm{Pl}\sim M_\mathrm{Pl}^2/M$ and a pressure $p=\rho$. It is in that sense that the dense black hole gas picture is akin to a stiff fluid.

To then get the viscosity (which has never been considered before for a dense black hole gas), let us estimate the interaction cross section by $\sigma_\mathrm{cs}\sim R^2$, where in analogy with the dilute gas of the previous subsection the propagation speed of fluctuations is essentially taken to be unity for a stiff fluid. The mean free path follows as $\ell_\mathrm{mfp}\sim 1/(n\sigma_\mathrm{cs})\sim R$ since $n=N/V\sim R^{-3}$. Already, we see that the mean free path is of the order of the Hubble radius by construction. Indeed, if black holes are expected to fill each Hubble patch, then it takes a distance $R\sim|H|^{-1}$ before black holes interact with one another. As such, we expect the na\"ive kinetic expression \eqref{eq:etakin} for viscosity to be only a rough order of magnitude estimate. Nevertheless, it should convey the right scaling as a function of energy density. The above mean free path implies $\eta\sim\rho R$, but the energy density is actually related to the black holes' radius as $\rho\sim(M_\mathrm{Pl}/R)^2$ as we saw above, hence we finally obtain
\begin{equation}
    \eta\sim\frac{M_\mathrm{Pl}^2}{R}\sim M_\mathrm{Pl}^2|H|\sim M_\mathrm{Pl}\sqrt{\rho}\,.\label{eq:etaBHG}
\end{equation}

It is interesting to notice that, from the results above, the ratio of viscosity to entropy density is constant (and of order unity), as was the case for the finite-temperature interacting scalar field.
As already mentioned, we might indeed expect the conjectured bound $\eta/s\geq 1/(4\pi)$ to hold.
In fact, by assuming the conjecture, the above result has already been guessed and consequences thereof explored in \cite{Masoumi:2014nfa}.
Another interesting observation is that \eqref{eq:etaBHG} implies $M_\mathrm{Pl}^2H^2\sim\rho$, which is the Friedmann constraint equation with no anisotropies. This is perhaps not a surprise since the derivation essentially assumes the universe to be isotropic enough for the dense black hole gas to form in the first place. However, it seems to suggest already that no anisotropy is allowed to form when the universe is dominated by such matter. As we saw from Sec.~\ref{sec:viscoanicosmo}, a viscosity coefficient of the form of \eqref{eq:etaBHG} does indeed lead to isotropisation, i.e., the energy density in anisotropies always remains subdominant compared to the energy density of a stiff fluid (the dense black hole gas in this case).
Therefore, this picture of a dense black hole gas represents the only microphysical origin known to the authors of a stiff fluid with viscosity given by the scaling $\eta\propto\rho^{1/2}$, which was previously phenomenologically understood to perfectly isotropise the universe \cite{Belinski:2017fas,Belinski:2013jua,Ganguly:2020daq}, i.e., leading to a Friedmann singularity if taken all the way to a big crunch.

\section{The evolution of anisotropies in various scenarios}\label{sec:evo}

In the previous section, we presented three fluids for which one can derive a viscosity coefficient from a microphysical perspective. The dilute and dense black hole gases can in fact be viewed as a single fluid in opposite limits, while the finite-temperature interacting scalar field is unambiguously different in nature. In deriving the properties of the black hole gas (in both the dilute and dense limits), we had to resort to the assumption that the background cosmology was isotropic to a good approximation in the first place. The question of how anisotropies (even if small initially) can evolve subsequently remains well posed.
The goal of this section is thus to explore the evolution of anisotropies for the fluids described in the previous section, first under the assumption of small anisotropies initially (which can apply to both black hole gases and the scalar field example), and then conversely, in the limit of large initial anisotropies (which can only be applied to the scalar field model).

\subsection{Small anisotropy limit}

For a black hole gas in the dense limit (where $\eta\propto\rho^{1/2}$), the evolution of anisotropies is already known from the analytical solution \eqref{eq:rhosigmaDBHG} and previous works \cite{Belinski:2017fas,Belinski:2013jua,Ganguly:2020daq} as already discussed, which confirms the isotropising power of such a fluid. In the dilute limit (where $\eta$ is constant), we also already obtained an analytical solution in \eqref{eq:rhosigmaconstanteta}, but a solution for the background scalar factor $a(t)$ is needed to fully quantify the evolution of anisotropies in such a case. This is where assuming small anisotropies initially (so approximately FLRW) can be useful analytically.

Under the assumption of a FLRW metric initially, we can solve for the evolution of a Bianchi-I metric for a wide class of viscous fluids. For the sake of generality, let us consider a phenomenological parametrisation of the viscosity coefficient as
\begin{equation}\label{eq:defeta}
    \eta=\kappa\left(\frac{\rho}{M_\mathrm{Pl}^4}\right)^nM_\mathrm{Pl}^3\,,
\end{equation}
where the constant $n$ determines how viscosity scales as a function of $\rho$ (e.g., $n=0$ for a dilute black hole gas, while $n=1/2$ for a dense black hole gas), and $\kappa\geq 0$ is the proportionality factor, whose exact value can be derived from the microphysics of the fluid.
In the above, we set up the dimensions such that $\kappa$ is a dimensionless constant this time.
With this parameterisation of viscosity, let us rewrite the background EOMs \eqref{eq:BIall} as follows,
\begin{subequations}
\begin{align}
    3M_\mathrm{Pl}^2H^2&=\rho\left(1+\Omega_\sigma\right)\,,\\
    2M_\mathrm{Pl}^2\dot H&=-\rho\left(1+w+2\Omega_\sigma\right)\,,\\
    \dot\rho+3H\rho(1+w)&=\kappa M_\mathrm{Pl}^{1-4n}\rho^{1+n}\Omega_\sigma\,,\\
    \partial_t\sigma_i{}^j+3H\sigma_i{}^j&=-\kappa M_\mathrm{Pl}^{1-4n}\rho^n\sigma_i{}^j\,,\label{eq:sigmamunurhon}
\end{align}
\end{subequations}
where $w:=p/\rho$ defines the matter EoS and where we defined
\begin{equation}
    \Omega_\sigma:=\frac{\rho_\sigma}{\rho}=\frac{M_\mathrm{Pl}^2\sigma^2}{\rho}
\end{equation}
to be the ratio of the shear energy density to the matter energy density, which we dub the shear-to-matter ratio.

The logic to solve the above analytically shall be the following: consider a contracting universe in which the energy density in anisotropies is initially contributing at most as much as the matter content, i.e., the ratio $\Omega_\sigma=\rho_\sigma/\rho$ is at most order 1 initially.
Then, one can say that, initially, $3H^2\simeq\rho/M_\mathrm{Pl}^2$ and $\dot\rho+3H(\rho+p)\simeq 0$ (provided $\kappa$ is also not too large) as a rough approximation.
In other words, one assumes that the spacetime is approximately FLRW at the onset of the analysis and check whether that approximation may remain valid under time evolution (i.e., whether it improves or worsens).
Practically speaking, this means checking whether or not $\Omega_\sigma$ remains $\leq 1$.
With no viscosity, we already saw that $\rho_\sigma$ grows as $a^{-6}$, so even if we start out with small anisotropies, the contraction will cause these anisotropies to grow to such an extent that the universe quickly becomes anisotropy dominated, certainly more than allowed for a successful bounce to occur or for a structure formation scenario to be realised.
We are now asking the question whether the inclusion of shear viscosity from a fluid that can reasonably be expected to be present can mitigate the growth of these anisotropies.
Our question is thus whether $\rho_\sigma$ may remain subdominant, and in fact, how much it may decay as the universe contracts under the influence of shear viscosity.

The solution to the EOM for $\sigma_i{}^j$, Eq.~\eqref{eq:sigmamunurhon}, reads
\begin{equation}
    \sigma_i{}^j(t)=\sigma_i{}^j(t_\mathrm{i})\,\mathrm{exp}\left[-\int_{t_\mathrm{i}}^t\mathrm{d}\tilde t\,\Big(3H(\tilde t)+\kappa M_\mathrm{Pl}^{1-4n}\rho(\tilde t)^n\Big)\right]\,,\label{eq:sigmaint}
\end{equation}
where $t_\mathrm{i}$ is the time at which the initial conditions are set.
Given the approximate FLRW background, we have
\begin{equation}
    H(t)\simeq\frac{2}{3(1+w)t}\,,\qquad\rho(t)\simeq\frac{4M_\mathrm{Pl}^2}{3(1+w)^2t^2}\,,
\end{equation}
where we shall be looking at the regime where $t<0$ for a period of contraction.
Performing the integral in \eqref{eq:sigmaint}, it follows that
\begin{align}
    &\Omega_\sigma(t)=\Omega_\sigma(t_\mathrm{i})\left(\frac{t_\mathrm{i}}{t}\right)^{\frac{2(1-w)}{1+w}}\nonumber\\
    &~\times\exp\left[-\frac{2^{1+2n}\kappa (M_\mathrm{Pl}|t_\mathrm{i}|)^{1-2n}}{3^n(1-2n)(1+w)^{2n}}\left(1-\left(\frac{t}{t_\mathrm{i}}\right)^{1-2n}\right)\right]\,,\label{eq:rhosigmarhogen}
\end{align}
as long as $n\neq 1/2$ (one has to treat the $n=1/2$ case separately) and where we used the fact that we have $t^{1-2n}=t(t^2)^{-n}<0$ for $t<0$, hence $t^{1-2n}=-|t|^{1-2n}$.

A first thing to notice from \eqref{eq:rhosigmarhogen} is that with no viscosity ($\kappa=0$), one is left with $\Omega_\sigma\propto|t|^{-2(1-w)/(1+w)}$, and therefore, one recovers the usual result that anisotropies grow, are constant, or decay compared to the background energy density as $t\rightarrow 0^-$ if $w<1$, $w=1$, or $w>1$, respectively (assuming the fluid's EoS parameter is always at least greater than $-1$).
Then, reinserting viscosity with $\kappa>0$, we note that if $n>1/2$,
the term in the exponential becomes dominated by $-(2n-1)(t_\mathrm{i}/t)^{2n-1}$, which goes to $-\infty$ as $t\to 0^-$.
Consequently, anisotropies are (exponentially) infinitely suppressed, and the BKL instability is resolved in this regime.
However, we do not know of a realistic fluid, which would have a well-defined viscosity all the way to high energy scales with $n>1/2$.\footnote{In fact, if a fluid has viscosity satisfying the relation $\eta\propto\rho^n$ with $n>1/2$ such that it fully isotropises the BKL singularity, it has been shown that it would necessarily imply superluminal propagation of viscous excitations \cite{Belinskii:1979,Belinski:2013jua,Belinski:2017fas}.}

The case $n=1/2$ is treated separately later, so let us focus on the cases when $n<1/2$. One can see from \eqref{eq:rhosigmarhogen} that as $t\to 0^-$, the factor in the exponential only goes to a finite negative constant, so while the anisotropies are exponentially suppressed compared to the non-viscous solution, the approach to the singularity remains highly anisotropic for $w<1$. The exponential suppression remains interesting though, especially in a context where the universe might not reach a singularity or even Planckian scales. Indeed, it might be interesting to see if there could be significant isotropisation before a bounce occurs. To explore this question, let us first observe that demanding the time derivative of \eqref{eq:rhosigmarhogen} to be negative at the initial time $t_\mathrm{i}$, we find that the shear-to-matter ratio $\Omega_\sigma$ is initially decaying (demanding $\dot\Omega_\sigma(t_\mathrm{i})<0$, so the universe is initially isotropising) as long as
\begin{equation}
    \kappa>\frac{3^{1-n}}{2}(1-w)\left(\frac{|H_\mathrm{i}|}{M_\mathrm{Pl}}\right)^{1-2n}=:\kappa_\mathrm{min}\,,\label{eq:kappamin}
\end{equation}
assuming $w>-1$, and where $H_\mathrm{i}=2/(3(1+w)t_\mathrm{i})$ is the initial value of the Hubble parameter.
What this shows is that, for $w=1$, any positive non-zero viscosity coefficient suffices to start isotropisation, i.e., for the shear-to-matter ratio $\Omega_\sigma$ to start decreasing.
When $w<1$, however, the viscosity coefficient cannot be arbitrarily small for that matter; it needs to be larger than a minimal value dubbed $\kappa_\mathrm{min}$. The smaller the EoS parameter, the larger $\kappa$ needs to be. Also, the higher the initial energy scale, the larger the viscosity coefficient must be to be able to begin isotropisation. Vice versa, when the universe is initially very large (small initial energy scale), the viscosity coefficient can be smaller.
This dependence on the initial Hubble scale is most important the closer $n$ is to $0$, but it becomes less important for $n$ closer to $1/2$.

Provided isotropisation starts, we can then check at what point there is a turnaround, i.e., a point where the shear-to-matter ratio $\Omega_\sigma$ starts growing again. By solving $\dot\Omega_\sigma=0$, we find that this occurs at an energy scale
\begin{equation}
    \frac{|H_\star|}{M_\mathrm{Pl}}=\left(\frac{2\kappa}{3^{1-n}(1-w)}\right)^{\frac{1}{1-2n}}\,.\label{eq:endofisoscale}
\end{equation}
This expression only applies for $w<1$; for $w\geq 1$, $\Omega_\sigma$ always decreases until a big crunch or a bounce is reached.
In fact, the larger $w$ is, the higher the energy scale at which isotropisation stops can be.
The same applies for the viscosity coefficient, as expected.
Additionally, the closer $n$ is to the value $1/2$, the more efficient isotropisation is.

In the cases where isotropisation stops, $\Omega_\sigma$ starts growing again, and one can approximate its subsequent evolution by the usual power-law scaling without viscosity,
\begin{equation}
    \Omega_\sigma(t)\approx\Omega_{\sigma,\star}\left(\frac{t}{t_\star}\right)^{\frac{2(1-w)}{1+w}}\,,
\end{equation}
where $t_\star=2/(3(1+w)H_\star)$ is the end-of-isotropisation time following from \eqref{eq:endofisoscale}, and $\Omega_{\sigma,\star}:=\Omega_\sigma(t_\star)$ is the corresponding value of the shear-to-matter ratio at that time.
The time at which the ratio reaches unity,
$t_\mathrm{c}=t_\star\Omega_{\sigma,\star}^{(1+w)/(2(1-w))}$,
represents the moment when anisotropies start dominating again, and so the moment when the initial assumption breaks down and the above solutions do not apply anymore.
Past that point, we essentially expect the universe to reach its chaotic mixmaster behavior toward the big crunch.

\begin{figure}
    \centering
    \includegraphics[scale=0.6]{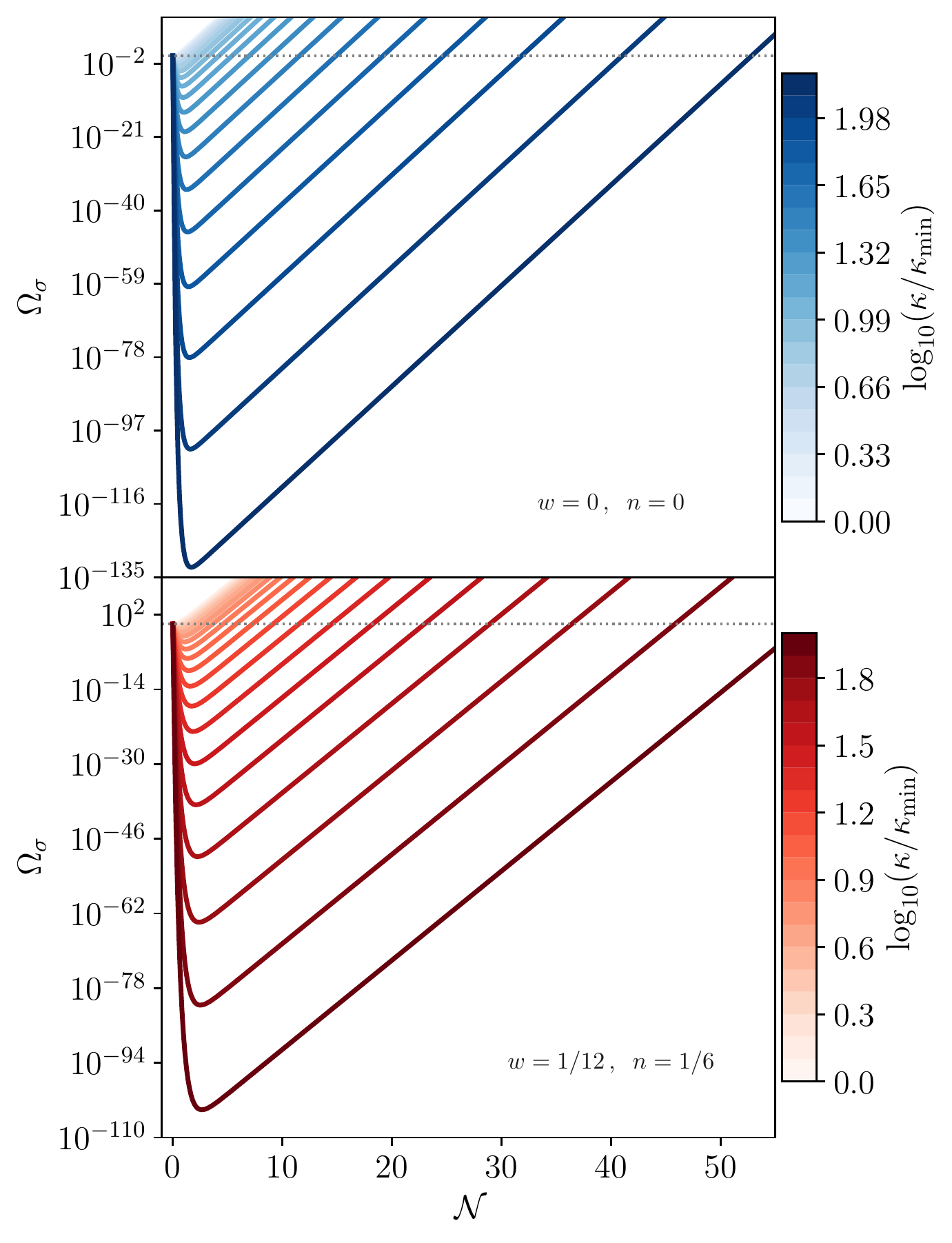}
    \caption{Plots of the shear-to-matter ratio $\Omega_\sigma=M_\mathrm{Pl}^2\sigma^2/\rho$ as a function of the $e$-folding number $\mathcal{N}\sim\ln(a|H|)$. The top plot shows the case of a dust-like EoS $w=0$ and constant viscosity coefficient ($n=0$), while the bottom plot shows an example for non-zero values with $w=1/12$ and $n=1/6$. The colors code as indicated by the color bars shows the value of the viscosity coefficient of proportionality $\kappa$ as a ratio of the minimal isotropising value $\kappa_\mathrm{min}$ derived in \eqref{eq:kappamin}. The initial conditions are set at a time $t_\mathrm{i}=-10^{80}\,t_\mathrm{Pl}$, and the initial shear-to-matter ratio is set to the threshold value $\Omega_{\sigma,\mathrm{i}}=1$. This value is highlighted by the horizontal dotted grey line.}
    \label{fig:smallani}
\end{figure}

Let us explore the above timescales in a specific model of interest.
Let us consider the case of a constant viscosity coefficient corresponding to $n=0$, which was already solved in \eqref{eq:rhosigmaconstanteta}. If we now assume the background to be approximately FLRW with matter having the EoS of dust ($w=0$), we can write this sub-case of \eqref{eq:rhosigmarhogen} as
\begin{equation}
    \frac{\Omega_\sigma}{\Omega_{\sigma,\mathrm{i}}}=\left(\frac{a_\mathrm{i}}{a}\right)^3\exp\left[-\frac{4\kappa}{3}\frac{M_\mathrm{Pl}}{|H_\mathrm{i}|}\left(1-\left(\frac{a}{a_\mathrm{i}}\right)^{3/2}\right)\right]\,.
\end{equation}
This is thus the solution for the evolution of anisotropies in the example of a contracting universe containing a dilute gas of black holes with effective EoS $w=0$, which is initially isotropic to a good approximation.
The evolution of $\Omega_\sigma$ in this case is shown in the top plot of Fig.~\ref{fig:smallani} as a function of the $e$-folding number defined according to
\begin{equation}
    \mathcal{N}:=\ln\left(\frac{aH}{a_\mathrm{i}H_\mathrm{i}}\right)\,.
\end{equation}
The bottom plot of Fig.~\ref{fig:smallani} shows similar computations, but applying \eqref{eq:rhosigmarhogen} for some phenomenological\footnote{Such arbitrary values could potentially correspond to some intermediate regime, in between a dilute and a dense black hole gas for instance, or in the case of the scalar field example, in between the matter- and radiation-dominated regimes.} case with $w=1/12$ and $n=1/6$.
Curves of different color show different values of the viscosity coefficient of proportionality $\kappa$, whose value as a fraction of the minimal isotropising coefficient $\kappa_\mathrm{min}$ can be read off from the color bar.
There, we can see that for $\kappa$ close to $\kappa_\mathrm{min}$ (the curves with lighter color), the universe does start by isotropising, but this is not very efficient, and $\Omega_\sigma$ quickly turns over and grows as a power law beyond $\Omega_\sigma=1$, indicating a future shear-dominated universe. It is only for values of $\kappa$ that are about 1 or 2 orders of magnitude larger than $\kappa_\mathrm{min}$ that we start seeing long-lasting isotropisation (of the order of tens of $e$-folds). In those cases (darker curves), isotropisation is extremely efficient (exponential as expected) for the first few $e$-folds before turnaround and power-law growth. However, since $\Omega_\sigma$ shrinks to exponentially small values at first, it takes several tens of $e$-folds before shear becomes dominant again.

The problem with the above description in the case of the physically motivated dilute black hole gas is that large viscosity coefficients $\kappa/\kappa_\mathrm{min}\sim\mathcal{O}(10^2)$ are not expected to respect previously discussed approximations.
To see this, let the constant viscosity coefficient $\eta=\kappa M_\mathrm{Pl}^3$ be given according to \eqref{eq:etafindBHG} for a dilute black hole gas. Together with \eqref{eq:kappamin} when $w=n=0$, we thus find
\begin{equation}
    \frac{\kappa}{\kappa_\mathrm{min}}\sim\frac{c_\mathrm{s}^5}{R|H_\mathrm{i}|}\,,
\end{equation}
which needs to be at the very least greater than $1$ for isotropisation to work, i.e., one needs $c_\mathrm{s}^5>R|H_\mathrm{i}|$. However, this is clearly incompatible with the requirement that the mean free path has to be smaller than the Hubble radius, cf.~\eqref{eq:smallcsconstr}. Therefore, we conclude that a dilute black hole gas is not viscous enough for isotropisation to start, even less so for an isotropic background to be sustained.

The evolution of anisotropies in the case of an interacting scalar field at high temperature in the small anisotropy limit was already found in Sec.~\ref{sec:viscoanicosmo}. Indeed, assuming the background to be FLRW and radiation dominated and taking the viscosity coefficient to be $\eta\propto a^{-3}$ in accordance with \eqref{eq:etafiniteT}, one finds the solution \eqref{eq:shearsoletaam3}, which depicts isotropisation as $a\to 0$. This solution is equivalent to \eqref{eq:rhosigmarhogen} with $w=1/3$ and $n=3/4$. As mentioned earlier, $n>1/2$ immediately implies isotropisation in this limit, but approximations most likely break down before reaching a singularity in this case. For this reason, this requires greater scrutiny, and so we defer the analysis of this scenario to the next subsection, where we look at the large anisotropy limit numerically, making no approximation about the background.

To end this subsection, we come back to the special case of $n=1/2$. In such a case, the integral \eqref{eq:sigmaint} yields
\begin{equation}
    \Omega_\sigma(t)=\Omega_{\sigma,\mathrm{i}}\left(\frac{t}{t_\mathrm{i}}\right)^{\frac{2}{1+w}\left(\frac{2}{\sqrt{3}}\kappa-(1-w)\right)}\,.
\end{equation}
Isotropisation thus occurs as $t\to 0$ only if
\begin{equation}
    \kappa>\frac{\sqrt{3}}{2}(1-w)\,.
\end{equation}
In the case of a stiff fluid with $w=1$, we see that any non-zero positive viscosity coefficient of proportionality leads to isotropisation, in accordance with the expectations previously mentioned.
We note that for a general EoS such a lower bound on the viscosity coefficient of proportionality has already been derived in \cite{Ganguly:2020daq}.
In fact, there it is found that for $n=1/2$, whenever
\begin{equation}
    \kappa>3(1-w)\,,
\end{equation}
the future crunching singularity is a stable Friedmann singularity (i.e., the universe fully isotropises by then). This has been derived for all Bianchi classes, and thus, it may explain the more stringent proportionality factor of 3 compared to $\sqrt{3}/2$ found in our simplified Bianchi-I analysis under the assumption of small anisotropies.

\subsection{Large anisotropy limit}

\subsubsection{Bianchi I}

As mentioned in the previous subsection, the case of an interacting scalar field at finite temperature has strong potential isotropising power, although this remained at the level of assuming small anisotropies initially. The strength of this model, though, lies in the fact that one does not have to make any assumption about the `formation' of the fluid or its previous history. In other words, even if the universe is highly anisotropic to start with, we would still reasonably expect a $\lambda\phi^4$ scalar field in a thermal bath to exhibit viscosity, and consequently, given its contribution to the coupled Einstein equations, affect the subsequent evolution of the anisotropies. This would even be true if one started in a maximally anisotropic homogeneous flat universe, also known as a Kasner universe. We shall be interested in this initial limit in this subsection, i.e., when anisotropies are dominant over everything else.

Let us first restrict ourselves to the case of flat spatial sections, i.e., to a Bianchi type-I metric (the case with curvature anisotropy, Bianchi IX, is treated separately later). We now seek to solve the corresponding equations \eqref{eq:BIall} numerically, where in the case of the field theory model introduced in Sec.~\ref{subsec:finite-temp-intro}, we can use \eqref{eq:etafiniteT} for the viscosity coefficient together with the usual scaling of temperature $T\propto 1/a$. In this case, the matter and shear EOMs \eqref{eq:mattershearEOMs} reduce to
\begin{subequations}\label{eq:EOMsfiniteT}
\begin{align}
    \dot\rho+4\frac{\dot a}{a}\rho&=\frac{4\alpha T_0^3}{\lambda^2}\left(\frac{a_0}{a}\right)^3\sigma^2\,,\\
    \partial_t\sigma_i{}^j+3\frac{\dot a}{a}\sigma_i{}^j&=-\frac{2\alpha T_0^3}{\lambda^2M_\mathrm{Pl}^2}\left(\frac{a_0}{a}\right)^3\sigma_i{}^j\,,
\end{align}
\end{subequations}
assuming the matter EoS $w=1/3$ for the radiation bath, and where we denote the (expected order 1) constant of proportionality in the viscosity coefficient \eqref{eq:etafiniteT} by $\alpha$. For the purpose of the numerical analysis, we simply set $\alpha=1$, and the scalar field self-interaction coupling constant is taken to be $\lambda=10^{-3}$. Other numerical values have been explored, but we focus our attention here on the free parameter $T_0$, which sets the initial temperature of the thermal bath at the initial scale factor value $a_0$. Exploring a range of values for $T_0$ shall encapsulate different choices for the combination of parameters $\alpha T_0/\lambda^2$.

\begin{figure}
    \centering
    \includegraphics[scale=0.6]{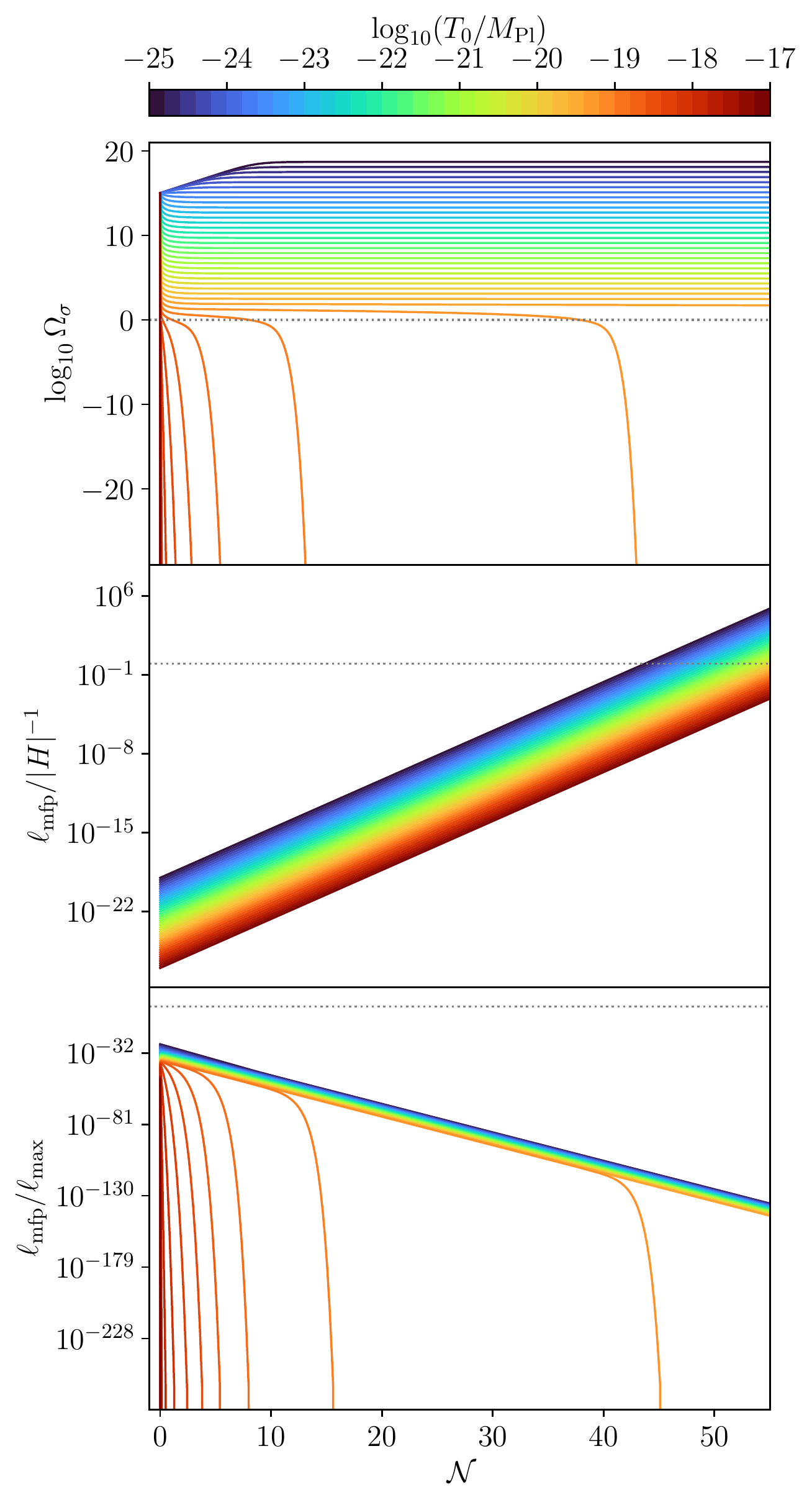}
    \caption{Plots of the shear-to-matter ratio (top plot), ratio of the mean free path over the Hubble radius (middle plot) and ratio of the mean free path over its maximal allowed value (bottom plot) as functions of the $e$-folding number $\mathcal{N}$. The curves of different color show different choices for the initial temperature $T_0$, as shown by the top color bar. The horizontal dotted grey line always indicates where the ratios cross unity. Successful isotropisation (with all approximations under control) is achieved when the curves are under this line.}
    \label{fig:BIfiniteT}
\end{figure}

We are now in position to numerically solve the set of coupled ordinary differential equations \eqref{eq:BIHEOM} and \eqref{eq:EOMsfiniteT}, which respect the constraint \eqref{eq:BIconstrgen}. Solutions are shown in Fig.~\ref{fig:BIfiniteT} for the shear-to-matter ratio $\Omega_\sigma$ (in the top plot) as a function of the $e$-folding number $\mathcal{N}$. Initial conditions are picked at the Hubble scale $H_0=-10^{-50}\,M_\mathrm{Pl}$ such that the initial shear-to-matter ratio is $\Omega_{\sigma,0}=10^{15}$, i.e., we want the anisotropies to be dominant over the matter at the initial time and see how this changes under time evolution. The curves of different color show different values of the initial thermal bath temperature $T_0$, ranging from colder ($10^{-25}\,M_\mathrm{Pl}$, blue) to warmer ($10^{-17}\,M_\mathrm{Pl}$, red).

Starting from the colder temperatures in blue in Fig.~\ref{fig:BIfiniteT}, we see that $\Omega_\sigma$ first grows as the usual power law in a Kasner universe, before starting to saturate. In fact, after about 10 $e$-folds, $\Omega_\sigma$ reaches a constant, already showing that viscosity has started becoming effective in mitigating the otherwise unbounded growth of anisotropies. For the lighter shades of blue and the green/yellow curves ($T_0\sim\mathcal{O}(10^{-23}-10^{-20})\,M_\mathrm{Pl}$), we can see that $\Omega_\sigma$ starts by decreasing, demonstrating isotropisation, but the exponential damping does not last. Rather, $\Omega_\sigma$ saturates at some constant value greater than $1$, meaning that the universe remains anisotropy dominated (though with bounded shear).

The situation changes once we consider initial temperatures warmer than about $10^{-19.5}\,M_\mathrm{Pl}$. For the darker orange curves, we see that there is initial isotropisation followed by saturation, but then there is a second phase of exponential isotropisation, which brings $\Omega_\sigma$ to exponentially small values, well below unity, such that the universe is isotropic to a very good approximation. For the red curves, this isotropisation occurs all at once, with no intermediate saturation phase, and the universe becomes isotropic within a few $e$-folds (or even a fraction of an $e$-fold for the temperatures closer to $10^{-17}\,M_\mathrm{Pl}$ and above).

In all of these cases, it is important to consider whether the viscosity approximation is valid though (and whether thermal equilibrium holds). In its simplest iteration, this can be stated as the situation when the mean free path $\ell_\mathrm{mfp}$ remains less than the characteristic length scale of the system. In our case, the characteristic length scale is given by the size of the horizon $|H|^{-1}$, where $H$ is the average expansion rate as before. For this reason, we show the ratio $\ell_\mathrm{mfp}/|H|^{-1}$ in the middle plot of Fig.~\ref{fig:BIfiniteT}. For our purposes, we use the expression for the mean free path in this field theory, derived in Eq.~\eqref{eq:mfp_def}, with the constant of proportionality set to $1$. We see that in all cases considered the approximation remains valid for at least 40 $e$-folds. For the higher initial temperatures that successfully lead to an isotropic universe, we see that the approximation remains valid even longer, up to at least $60$ $e$-folds. Once the mean free path becomes of the order of the Hubble radius and even surpasses it, the expression used for viscosity does not apply anymore. In fact, one would rather expect viscosity to go to zero as thermal equilibrium is lost in the limit where the averaged volume shrinks to zero. Therefore, one cannot realistically expect isotropisation to remain effective all the way to a crunching singularity. Rather, a Kasner singularity is anticipated. Yet, considering the efficiency of the exponential damping of shear within the regime of validity of the theory initially when $T_0$ is high enough, even if one were to turn off viscosity altogether once $\ell_\mathrm{mfp}\sim|H|^{-1}$, it would take several hundreds of $e$-folds (if not more) before the power-law growth in anisotropies would bring $\Omega_\sigma$ back to values greater than unity. Therefore, it is expected that in any realistic scenario where the universe would undergo a non-singular bounce at some high-curvature scale such that a singularity is never reached, the universe would still be highly isotropic at the onset of the transition from contraction to expansion.

At last, let us point out that according to \eqref{eq:validitymfp} another approximation should be satisfied, namely $\ell_\mathrm{mfp}/\ell_\mathrm{max}\lesssim 1$, where $\ell_\mathrm{max}$ is defined in Eq.~\eqref{eq:deflmax}. This can be computed using the usual radiation entropy relation $s\propto T^3$. The result is shown in the bottom plot of Fig.~\ref{fig:BIfiniteT}, where it can be seen that the ratio $\ell_\mathrm{mfp}/\ell_\mathrm{max}$ remains well below unity throughout the evolution and for any initial temperature in the given range. In fact, the approximation improves under time evolution and for warmer initial temperatures. Therefore, the requirement that the Maxwell relaxation time be small enough does not represent a threat to the validity of the viscosity approximations in this context.

\subsubsection{Bianchi IX}

The above discussion only concerns expansion anisotropies, where the underlying geometry is a flat anisotropic universe. However, the generic approach to a singularity and, in our case, the endpoint of contraction is the closed anisotropic universe described by the Bianchi type-IX metric (see, e.g., \cite{Kiefer:2018uyv} and references therein). In this case, the anisotropy energy density is not just stored in the expansion tensor, but there is also an anisotropy `potential'. This potential is nothing but the anisotropic $3$-curvature terms that arise in the closed Bianchi type-IX universe, and which are responsible for the chaotic mixmaster oscillations on approach to a singularity.

The Bianchi-IX metric takes the general form of a homogeneous spacetime as follows,
\begin{equation}
    g_{ab}\mathbf{d}x^a\otimes\mathbf{d}x^b=-\mathbf{d}t\otimes\mathbf{d}t+h_{ij}\bm{\sigma}^i\otimes\bm{\sigma}^j\,.
\end{equation}
Here, $h_{ij}$ is the spatial metric, and the $\bm{\sigma}^i$'s are one-forms, which take the simple Cartesian form $\mathbf{d}x^i$ in the case of a flat anisotropic universe [Bianchi I, cf.~\eqref{eq:BIMetric}]. In the case of homogeneous spacetimes with non-trivial curvature, they can always be chosen so that $h_{ij}$ always remains strictly a function of time. In the case of the Bianchi-IX universe, these one-forms take the following shape,
\begin{align}
    \bm{\sigma}^1&=-\sin\psi\,\mathbf{d}\theta+\cos\psi\sin\theta\,\mathbf{d}\varphi\,,\nonumber\\
    \bm{\sigma}^2&=\,\cos\psi\,\mathbf{d}\theta+\sin\psi\sin\theta\,\mathbf{d}\varphi\,\nonumber\\
    \bm{\sigma}^3&=\,\cos\theta\,\mathbf{d}\varphi+\mathbf{d}\psi\,,
\end{align}
which are the differential forms on a 3-sphere with coordinate ranges $0\leq\theta\leq\pi$, $0\leq\varphi\leq 2\pi$, and $0\leq\psi\leq 4\pi$. In the frame in which the metric $h_{ij}$ is diagonal and strictly a function of time, it takes the form
\begin{equation}
    h^i{}_j=a^2\,\mathrm{diag}\left(e^{2\beta_++2\sqrt{3}\beta_-},\,e^{2\beta_+-2\sqrt{3}\beta_-},\,e^{-4\beta_+}\right)\,.
\end{equation}
The volume averaged expansion is given by the scale factor $a(t)$. The variables $\beta_\pm(t)$ are the Misner variables that are used to parameterise the anisotropies.
The shear anisotropy has only two independent components as the anisotropic shear is traceless. In this formalism, the three-dimensional curvature ${}^{(3)}\!R$ on spatial hypersurfaces of constant coordinate time is given by
\begin{equation}
    {}^{(3)}\!R=-\frac{2}{a^2}U(\beta_+,\beta_-)\,,
\end{equation}
where the curvature potential $U(\beta_+,\beta_-)$ is given by
\begin{align}
    U(\beta_+,\beta_-)=&~\frac{1}{4}e^{-8\beta_+}-e^{-2\beta_+}\cosh\left(2\sqrt{3}\beta_-\right)\nonumber\\
    &+e^{4\beta_+}\sinh^2\left(2\sqrt{3}\beta_-\right)\,.
\end{align}
The Einstein equations \eqref{eq:EFE-orthonormal} in this formalism become (we set $M_\mathrm{Pl}=1$ for the rest of this subsection)
\begin{subequations}\label{eq:BIXeveq}
\begin{align}
    &3H^2=\rho+3\left(\dot\beta_+^2+\dot\beta_-^2\right)+\frac{1}{a^2}U(\beta_+,\beta_-)\,,\label{eq:constrBIX}\\
    &-2\dot H=\rho+p+6\left(\dot\beta_+^2+\dot\beta_-^2\right)+\frac{2}{3a^2}U(\beta_+,\beta_-)\,,\\
    &\dot\rho+3H(\rho+p)=12\eta\left(\dot\beta_+^2+\dot\beta_-^2\right)\,,\\
    &\ddot\beta_\pm+3H\dot\beta_\pm+\frac{1}{6a^2}\partial_{\beta_\pm}U=-2\eta\dot\beta_\pm\,,
\end{align}
\end{subequations}
where the shear energy density is $\sigma^2=3(\dot\beta_+^2+\dot\beta_-^2)$.

\begin{figure*}
    \centering
    \includegraphics[scale=0.6]{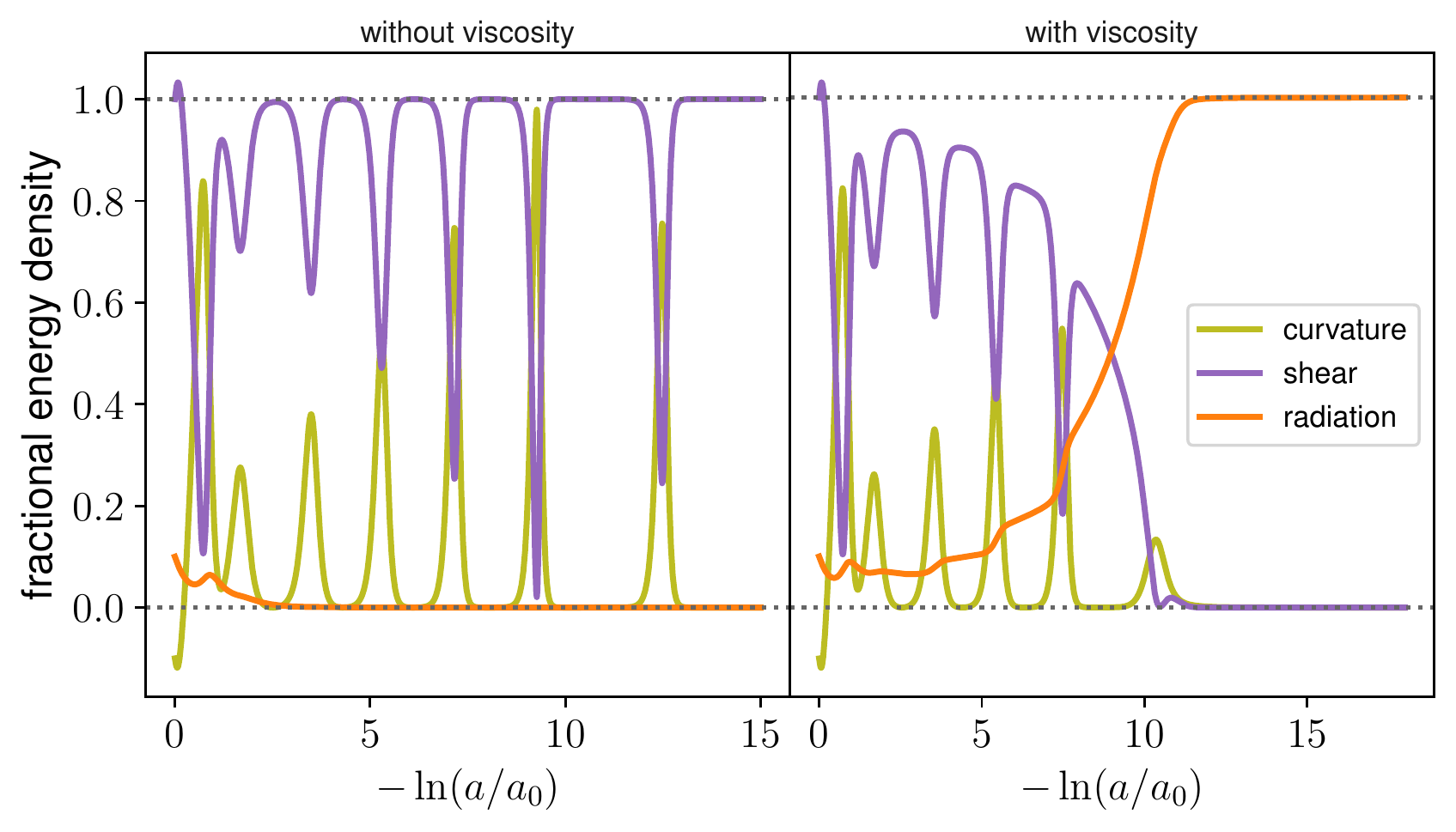}
    \caption{Plot of the fractional energy density ($X/3H^2$) for the different contributions $X$ to the total energy budget under forward time evolution [as a function of $N=-\ln(a/a_0)$]. The different contributions are: $X=\rho$ (matter [radiation] in orange), $X=\sigma^2$ (shear in purple), and $X=-{}^{(3)}\!R/2=U/a^2$ (curvature in olive). The sum of the three contributions always adds up to $3H^2$ in accordance with the constraint equation \eqref{eq:constrBIX}, hence the total fractional energy density is $1$, as depicted by the top horizontal dotted grey line. The bottom horizontal dotted grey line at $0$ indicates exponentially small contribution. The left plot shows the standard evolution without viscosity, while the right plot shown an example once viscosity is taken into account.}
    \label{fig:BIXfiniteT}
\end{figure*}

If one ignores the presence of viscosity and simply set $\eta\equiv 0$, then one recovers the usual chaotic mixmaster behaviour in the approach to a singularity. To see this, let us numerically solve the above set of ordinary differential equations when the matter content is radiation-like ($p=\rho/3$). The initial conditions are set in a contracting phase with $H_0=-10^{-40}$, $\beta_{+,0}=10^{-10}$, $\beta_{-,0}=-10^{-1}$, $\beta_{+,0}'=-1$, $\beta_{-,0}'=0$, and $a_0\approx 9.1\times 10^{39}$, where a prime here denotes a derivative with respect to the $e$-folding number $N:=-\ln(a/a_0)$, which turns out to be an easier time variable\footnote{The $e$-folding numbers $\mathcal{N}$ and $N$ only differ by a factor of $(1+3w)/2$ for a power-law solution $a(t)\propto|t|^{2/(3(1+w))}$. In particular, they are equal when $w=1/3$, and $\mathcal{N}$ ticks twice as fast as $N$ when $w=1$.} to work with in Bianchi IX, numerically speaking. Such values are chosen such that, initially, $\rho_0/(3H_0^2)=1/10$, $\sigma_0^2/(3H_0^2)=1$, and $-{}^{(3)}\!R_0/(6H_0^2)=-1/10$. Physically, this means that we choose shear anisotropies to be dominant over matter (radiation) initially since $\sigma_0^2/\rho_0=10$, but we want curvature anisotropies to be small. Taking $|\beta_\pm|\ll 1$ initially, the anisotropy potential is negative (the potential minimum is $-3/4$), indicating positive spatial curvature (${}^{(3)}\!R>0$), and the curvature radius is set to be small by taking $a_0$ large. In other words, we want to start in a large universe relatively close to a flat Bianchi-I spacetime in this example.

The result of the evolution is shown in the left plot of Fig.~\ref{fig:BIXfiniteT}. There, we see that, without viscosity, the radiation contribution (orange curve) rapidly goes to $0$, while anisotropies dominate. In fact, there is a chaotically oscillatory exchange between shear anisotropies (purple curve) and curvature anisotropies (olive curve), representative of the mixmaster dynamics as the universe approaches a BKL singularity.

When viscosity is introduced, the situation changes, as can be seen in the right plot of Fig.~\ref{fig:BIXfiniteT}. There, we numerically solve the same previous set of equations with the same initial conditions, except now the viscosity coefficient is taken to be $\eta=\alpha T_0^3(a_0/a)^3/\lambda^2$ as in the previous subsection. Numerical values for this example are taken to be $\alpha=1$, $\lambda=10^{-3}$, and $T_0=10^{-16}$.

For the first $e$-fold or so, the evolution with and without viscosity is very similar. However, as the scale factor decreases, the temperature rises and so does the viscosity coefficient. Accordingly, the radiation component is not diluted with respect to the anisotropies; rather, it remains more or less constant and starts growing after a few $e$-folds. Counterbalancing, the contribution from shear starts decreasing already after about $2$ $e$-folds. By $\approx 8.98$ $e$-folds, radiation becomes dominant over shear and curvature anisotropies ($\Omega_\sigma$ becomes smaller than unity). From then on, the spacetime becomes more and more isotropic, with anisotropies decaying to exponentially small values, and with the chaotic oscillations in the anisotropies stopping. Through this evolution, the mean free path $\ell_\mathrm{mfp}$ is found to remain smaller than the Hubble radius, up to approximately $35.66$ $e$-folds. By then, $\log_{10}\Omega_\sigma\approx -67.71$. As discussed in the previous subsection, beyond this point one cannot fully trust the approximations leading to the viscosity coefficient, which should in fact start decreasing. Nevertheless, even if viscosity were to suddenly become negligible again, it would take more than about $78$ $e$-folds before anisotropies would become dominant again. Therefore, we can say that the model is isotropic to a very good approximation for more than $100$ $e$-folds in total, and the warmer the initial temperature $T_0$ of the thermal bath, the longer the isotropic phase, in the same spirit as seen in Fig.~\ref{fig:BIfiniteT} for Bianchi I. In the end, it seems that isotropisation due to viscosity in a finite-temperature field theory is robust against curvature anisotropies, i.e., the same qualitative results hold whether the spacetime is of Bianchi type I or IX.

\section{Implications for gravitational waves}\label{sec:GWs}

So far, we have discussed the process of isotropisation in a contracting universe. We have shown that the addition of shear viscous anisotropic stress leads to a reduction of the fractional contribution of the shear anisotropies in many instances. The shear anisotropies that we have studied so far have been in spatially homogeneous cosmological settings --- they are non-perturbative by definition. Then, the perturbative limit of this represents a homogeneous and isotropic universe, but containing gravitational wave perturbations. The concept of an isotropic universe sourced by gravitational waves being equivalent to an anisotropic universe is not a new one. For example, in an open or flat isotropic Friedmann model, gravitational waves superimposed upon the background only leave homogeneity untouched, and hence reproduce the corresponding spatially homogeneous, anisotropic cosmology when the wavelength is infinitely long \cite{lukashGW}. This fulfills the assumption of homogeneity as the periodicity of a propagating wave with finite wavelength would actively violate it. There are some exceptions to this rule, such as in the case of circularly polarised gravitational waves \cite{lukashGW}, where the average quantities coincide with the scenario of a gravitational tensor representing a homogeneous isotropic cosmological model being sourced by gravitational wave anisotropies. The approach to a singularity also becomes quasi-isotropic and resembles the Friedmann solution. In general, one can assume that whatever physical effect modifies the propagation of shear will also correspondingly affect the propagation of gravitational waves (and vice versa). Examples include massive gravity \cite{Lin:2017fec}, neutrinos and more (e.g., \cite{Weinberg:2003ur,Pritchard:2004qp,Watanabe:2006qe,Stefanek:2012hj,Dent:2013asa,Baym:2017xvh,Kite:2021yoe,Brevik:2019yma,Goswami:2016tsu,Lu:2018smr}).

As there appears to be an inexorable link between the shear anisotropies we have been studying and gravitational waves, it would be interesting to see how a characteristic spectrum of gravitational waves would be affected at the end of shear viscosity driven contraction. For the purposes of this computation, we shall restrict ourselves to a flat background. This indicates the case of Bianchi I. In fact, we can even assume the background to be flat FLRW as the only anisotropies present are in expansion and can be written as part of the energy density. This is a very simple example of the general idea that perturbative shear anisotropies on a homogeneous background can be represented as gravitational waves on an isotropic background.

The general equations of motion in a homogeneous background given in \eqref{eq:EFE-orthonormal}, written in terms of the electric and magnetic parts of the Weyl curvature tensor denoted by $E_{ab}$ and $H_{ab}$, are given by \cite{ellis_maartens_maccallum_2012}
\begin{subequations}
\begin{align}
    \dot{\sigma}_{ab}+2H\sigma_{ab}+E_{ab}=&~\frac{1}{2M_\mathrm{Pl}^2}\pi_{ab}\,,\label{eq:dotsigmaEB}\\
    \dot{E}_{ab}+3HE_{ab}-\mathrm{curl}~H_{ab}=&-\frac{1}{2M_\mathrm{Pl}^2}\Big[(\rho+p)\sigma_{ab}\nonumber\\
    &\qquad+\dot{\pi}_{ab}+H\pi_{ab}\Big]\,,\label{eq:dotEEB}\\
    \dot{H}_{ab}+3HH_{ab}+\mathrm{curl}~E_{ab}=&~\frac{1}{2M_\mathrm{Pl}^2}\mathrm{curl}~\pi_{ab}\,.
\end{align}
\end{subequations}
These equations are written assuming linear perturbations around a flat background and follow the covariant and gauge-invariant approach to perturbation theory outlined in \cite{ellis_maartens_maccallum_2012}. They are easily generalisable to the fully non-linear case as for example in \cite{ellis_maartens_maccallum_2012, 1992ApJ...395...34B}.
This is different from the metric perturbation approach where we linearise the metric around a background and then trace the time evolution of the metric perturbations through the perturbed Einstein equations. The disadvantage of this approach is of course that it is hard to generalise to non-linear perturbations. The relative advantage of the latter approach is that we start out from the full non-linear equations \eqref{eq:EFE-orthonormal} and then linearise around a given background, in our case it would be the FLRW background. Gravitational wave perturbations in the metric-perturbation approach are the tensor modes born out of perturbations to the $ij$ components of the metric tensor and then traced through the Einstein equations. In contrast, in the covariant, gauge-invariant approach, gravitational wave perturbations are expressed as curvature perturbations that propagate and manifest themselves in the evolution of the electric and magnetic parts of the Weyl tensor. Pure tensor modes must be transverse and tracefree and therefore cause the divergence of the electric and magnetic parts of the Weyl tensor to disappear, i.e., $\mathrm{D}_bE_a{}^b=0$ and $\mathrm{D}_bH_a{}^b=0$, as well as the divergences of the shear anisotropy tensor and the anisotropic stress to disappear, $\mathrm{D}_b \sigma_a{}^b=0$ and $\mathrm{D}_b \pi_a{}^b=0$.

In the linearised limit around FLRW and the presence of anisotropic stress of the form of \eqref{eq:defeta1}, one can take a time derivative of \eqref{eq:dotsigmaEB} and use \eqref{eq:dotEEB} to derive a wave equation for the shear anisotropies $\sigma_{ij}$ \cite{ellis_maartens_maccallum_2012}, reminiscent of the wave equation obeyed for gravitational waves,
\begin{align}
    &\ddot{\sigma}_{ij}+\left(5H+\frac{2\eta}{M_\mathrm{Pl}^2}\right)\dot{\sigma}_{ij}\nonumber\\
    &+\left(\frac{1}{M_\mathrm{Pl}^2}\Big((\rho-3p)+2(\dot\eta+2H\eta)\Big)-\frac{\partial^2}{a^2}\right)\sigma_{ij}=0\,.\label{eq:sigmawaveeq}
\end{align}
In fact, perturbing the spatial metric as
\begin{equation}
    g_{ij}=h_{ij}=a^2(\delta_{ij}+\gamma_{ij})\,,
\end{equation}
where $\gamma_{ij}$ is the transverse and traceless tensor perturbation corresponding to the gravitational wave perturbation, the shear anisotropy tensor is related to the metric tensor perturbation as follows (e.g., \cite{Pereira:2019mpp}),
\begin{equation}\label{eq:relhsig}
    \sigma_{ij}=\frac{1}{2}a^2 \partial_t\gamma_{ij}\,,\qquad\sigma_i{}^j=\frac{1}{2}\partial_t\gamma_i{}^j\,.
\end{equation}
This is because the shear tensor ultimately is the traceless part of the expansion tensor defined by \eqref{eq:defsheartensot}, which is related to the time derivative of the metric variables in a homoegeneous spacetime. In drawing the equivalence between the metric perturbation approach to perturbation theory and the covariant gauge invariant approach, this relation would allow us to recover the familiar evolution equation for the metric tensor modes $\gamma_{ij}$ through Eq.~\eqref{eq:dotsigmaEB}.\footnote{We can also see this by noting that the electric part of the Weyl tensor $E_{ab}$ is related to the traceless part of the $3$-Ricci tensor denoted by ${}^{(3)}\!R_{\langle ab \rangle}$ as
\begin{equation*}
    {}^{(3)}\!R_{\langle ab \rangle} =E_{ab} + \frac{1}{2M_\mathrm{Pl}^2}\pi_{ab} - H\sigma_{ab}+\sigma_{c\langle a}\sigma_{b\rangle}{}^c
\end{equation*}
This relation is taken to be in the absence of vorticity, as in all of this work. The full equations are found in the Appendix of \cite{ellis_maartens_maccallum_2012}.}
The corresponding equation is of the form
\begin{equation}
    \partial_t^2\gamma_i{}^j+\left(3H+\frac{2\eta}{M_\mathrm{Pl}^2}\right)\partial_t\gamma_i{}^j-\frac{\partial^2}{a^2}\gamma_i{}^j=0\,,\label{eq:gammawaveeq}
\end{equation}
agreeing with, e.g., \cite{Fanizza:2021ngq,Goswami:2016tsu}.
In the infrared limit, i.e., on large super-Hubble scales where $\partial^2/a^2\to 0$, the equation becomes
\begin{equation}
    \partial_t^2\gamma_i{}^j+\left(3H+\frac{2\eta}{M_\mathrm{Pl}^2}\right)\partial_t\gamma_i{}^j\simeq 0\,.\label{eq:gammawaveeqIR}
\end{equation}
This can also be found by substituting \eqref{eq:relhsig} into the previously derived equation \eqref{eq:sigmaijBI}, which makes the connection between anisotropies and long-wavelength gravitational waves explicit.

A key aspect of the above, either viewed through \eqref{eq:sigmawaveeq} or \eqref{eq:gammawaveeq}, is that shear and equivalently gravitational waves receive a damping factor (in the form of a friction term) due to the presence of viscosity with $\eta>0$. The negativity of the Hubble parameter in a contracting universe typically implies the growth of shear and of gravitational waves (most easily seen on super-Hubble scales).\footnote{This is a problem, for instance, in the context of matter bounce cosmology, where a scale-invariant power spectrum of tensor perturbations is amplified to the same extent as scalar perturbations, resulting in an order unity tensor-to-scalar ratio (see, e.g., \cite{Quintin:2015rta,Li:2016xjb,Lin:2017fec}).} The viscosity coefficient can counterbalance this effect though, such that anisotropies are damped (resulting in isotropisation) and so are gravitational waves. In fact, in the FLRW limit, one can solve \eqref{eq:gammawaveeqIR} for the long-wavelength $\partial_t\gamma_i{}^j$ in the same way we solved for $\sigma_i{}^j$ in \eqref{eq:sigmaint}, from which we can translate the results. For a constant viscosity coefficient and a pressureless EoS, one finds an exponential damping initially [in the form of \eqref{eq:rhosigmaconstanteta}, where we should think of $\rho_\sigma$ being replaced by $\rho_\mathrm{GW}:=(M_\mathrm{Pl}^2/8)\partial_t\gamma_i{}^j\partial_t\gamma_j{}^i$]. A similar result was derived in \cite{Hawking:1966qi} for a constant coefficient of viscosity. In the context of matter bounce cosmology, this damping would not realistically resolve the large tensor-to-scalar ratio problem if the viscosity is coming from a dilute gas of black holes (for the same reason it could not realistically lead to isotropisation within the regime of validity of the approximations). For an interacting field theory at finite temperature with $\eta\propto 1/a^3$ and a radiation EoS, one recovers exponential damping in the form of \eqref{eq:shearsoletaam3}. For a dense black hole gas with $\eta=\kappa |H|$ (when $H<0$) and a stiff EoS, one finds in a similar way to \eqref{eq:rhosigmaDBHG} that $\rho_\mathrm{GW}\propto 1/a^{2(3-2\kappa/M_\mathrm{Pl}^2)}$, and hence gravitational waves are completely damped out by the time $a\to 0$ if $\kappa>3M_\mathrm{Pl}^2/2$.

\section{Discussion and conclusions}\label{sec:conclusions}

Bouncing cosmologies present an alternative to traditional expanding cosmologies by avoiding an initial singularity. The expense occurs by hypothesising some possible new physics at the bounce, which causes the universe to re-expand after an initial phase of contraction. However, there are a few problems regarding the growth of anisotropies and inhomogeneities in the contracting phase itself. Traditionally, a phase of ekpyrosis, where a fast-rolling scalar field mediates a slow contraction, exhibits an effective EoS $p \gg \rho$ and is able to dominate over the anisotropies and inhomogeneities.

Other dissipative mechanisms, such as particle creation and other quantum effects (e.g., \cite{1972JETP...34.1159Z,1974JETP...39..742L,Hu:1978zd,Hartle:1980nn,Calzetta:1986ey}), a non-linear EoS (e.g., \cite{Bozza:2009jx,Ganguly:2019llh}), and the introduction of shear viscosity have been studied in the context of anisotropy reduction. In this work, we have studied possible microphysical realisations of such a dissipative model of shear viscosity. We have studied this in the context of a gas of black holes, both in the dilute and the dense limit. We find that the coefficient of viscosity remains constant and is temporarily effective in suppressing anisotropies in the dilute limit. However, the viscosity approximation is violated unless the viscosity coefficient is small enough, in which case isotropisation cannot occur. In the dense black hole gas case (which is considerably more speculative), we have the beginnings of a microphysical picture of understanding how a coefficient of viscosity that scales with energy density as $\eta \propto \rho^{1/2}$ can be realised and, as has been seen in the literature, give rise to successful isotropisation and lead to a Friedmann singularity (if allowed to evolve to a crunch) even in the most general of anisotropic spatially homogeneous universes.

Another microphysical example that we have studied is the case of a $\lambda \phi ^4$ interacting scalar field theory at finite temperature. The effective evolution of the background is that of a radiation-dominated universe. We studied the evolution of anisotropies in the case of a flat Bianchi type-I universe containing only expansion anisotropies, as well as in the case of a spatially curved closed anisotropic Bianchi type-IX universe. We found that in both cases the viscous damping dissipates the energy density in the anisotropy into radiation. The viscosity approximation itself remains valid in both cases, at least for enough $e$-folds for the exponential suppression of anisotropies to be effective, under assumptions of high initial temperature and a universe that does not start out curvature dominated deep in the contracting phase for the case of Bianchi IX. Similar results have been found in the same context, but using different analyses and in the context of particle creation and semi-classical gravity \cite{Calzetta:1986ey}. Finally, as the anisotropy tensor itself is related to the time derivative of the tensor modes, the effect of the shear dissipation is equivalent to a damping of the amplitude of long-wavelength gravitational waves (see, e.g., \cite{Loeb:2020lwa,Mottola:1985ee} for additional implications of this principle).

While the $\lambda\phi^4$ model is an interesting toy model, which successfully manifests isotropisation, it does not constitute a complete theory of the very early universe. In particular, it cannot explain the formation of structures, i.e., it does not generate a nearly scale-invariant spectrum of curvature perturbations on large scales by itself. The addition of a spectator field (e.g., \`a la curvaton \cite{Cai:2011zx}) could potentially resolve this issue, but this would require further investigation, especially with regard to the competition between quantum and thermal fluctuations in such a model. Alternatively, a contracting $\lambda\phi^4$ model could be part of a larger scenario that includes a period of inflation (e.g., \cite{Qiu:2015nha,Graham:2019bfu,Ji:2021mvg}), which takes care of generating the right perturbations.

For the matter bounce scenario, where scale-invariant curvature perturbations are generated during a phase of matter-dominated contraction, it appears viscosity can serve as an isotropising mechanism to keep the model close enough to FLRW. However, this remains phenomenological since viscosity is actually hard to generate in a fluid that weakly interacts by definition. For example, we showed in this paper that a dilute gas of black holes could not realistically provide sufficient viscosity to keep the universe isotropic. Thus, unless one modifies the gravitational theory, e.g., with a graviton mass \cite{Lin:2017fec}, which suppresses both anisotropies and gravitational waves, or with a specific non-minimal coupling to gravity (e.g., \cite{Nandi:2019xag,Nandi:2020sif,Nandi:2020szp}, but see also \cite{Akama:2019qeh}), the matter bounce scenario remains unviable.

In any more realistic bouncing scenario hoping to explain the origin of the cosmic microwave background, one has to be aware that requiring isotropy with $\Omega_\sigma<1$ for a certain number of $e$-folds might not be sufficient. Indeed, $\Omega_\sigma$ might have to be several orders of magnitude below unity for the bounce to be achievable and for cosmological perturbations not to receive significant contributions from the shear. This is due to the fact that shear enters as a source term in the scalar, vector, and tensor perturbations of an anisotropic universe such as Bianchi I (see, e.g., \cite{Pereira:2007yy}). Therefore, one expects an upper bound on the size that $\sigma^2$ may be allowed to reach in any given scenario \cite{Ed}.

Another aspect that needs to be taken into consideration in a more realistic scenario is the presence of shear due to quantum fluctuations in addition to the classical anisotropies discussed in this work. For instance, stochastic fluctuations of a scalar field could produce an anisotropic stress sourcing shear. However, when the background EoS satisfies $w\geq 0$ as studied in this work, the resulting quantum shear only becomes dominant near the Planck scale \cite{Grain:2020wro}. Therefore, any `low-energy' bounce could evade this issue, though it remains an important contribution to shear that needs to be considered seriously in light of the previous paragraph.

Let us end by commenting on the dense black hole gas. As we mentioned, this remains the only known model resulting in $\eta\propto\sqrt{\rho}$ and thus in full isotropisation within the approximations. Such a gas remains a fairly exotic toy model though. To start, the possible formation channels of such a gas remain hand-wavy; dealing with large inhomogeneities and their collapse into black holes would certainly have to be tackled numerically as in, e.g., \cite{Clifton:2017hvg,deJong:2021bbo}. Also, there is a great lack of understanding of the evolution of black holes embedded in cosmological backgrounds (apart from approximately Schwarzschild-de Sitter and McVittie spacetimes --- see, e.g., \cite{Bousso:1997wi,Gregory:2018ghc,Kaloper:2010ec,Faraoni:2012gz,Faraoni:2013aba}), and refining the corresponding approximations made on that front would definitely improve the description of the dense black hole gas. Nevertheless, if such a gas could really exist in nature in the very early universe (near a crunching singularity for instance), it remains interesting to ask the question of what could be the possible subsequent evolution of the gas. Could the black holes pass through a bounce and become primordial black holes as suggested in \cite{Carr:2011hv,Clifton:2017hvg,Carr:2017wkz,Coley:2020ykx} or evaporate into remnants accounting for dark matter \cite{Rovelli:2018hba,Rovelli:2018hbk,Barrau:2021spy}? Could the black holes become stringy in nature at high energies and be part of a greater string-cosmology scenario \cite{Veneziano:2003sz,Quintin:2018loc}? Or could the black holes evaporate and emit specific electromagnetic signals or merge and emit specific gravitational-wave signals \cite{Barrau:2017ukm,Papanikolaou:2020qtd}? All those questions deserve closer scrutiny and could open up the path to a new understanding of the physics near the highest cosmological energy scales.

\begin{acknowledgments}
The authors acknowledge the stimulating atmosphere at McGill University, Dartmouth College and Nordita while this project was initiated and prepared over the years and thank Robert Brandenberger for insightful discussions and encouragement to pursue this project in the first place. This project also progressed thanks to discussions following the program Physics of the Early Universe --- An Online Precursor (code: ICTS/peu2020/08) of the International Centre for Theoretical Sciences (ICTS). J.\,Q.\ further thanks the Department of Applied Mathematics and Theoretical Physics (DAMTP), University of Cambridge for kind hospitality while this work was prepared and Jean-Luc Lehners, Edward Wilson-Ewing, and Maurizio Gasperini for valuable discussions. C.\,G.\ would like to thank the Cambridge Philosophical Society for the Henslow Fellowship. They would also like to thank Wolfson College, Cambridge and DAMTP, University of Cambridge for hosting them for the duration of the fellowship. Through the completion of this work, research at the Albert Einstein Institute has been supported by the European Research Council (ERC) in the form of the ERC Consolidator Grant CoG 772295 ``Qosmology'', and J.\,Q.\ further acknowledges financial support in part from the \textit{Fond de recherche du Qu\'ebec --- Nature et technologies} postdoctoral research scholarship and the Natural Sciences and Engineering Research Council of Canada Postdoctoral Fellowship.
\end{acknowledgments}

\bibliographystyle{JHEP2}
\bibliography{references}

\providecommand{\url}[1]{#1}\providecommand{\href}[2]{#2}\begingroup\raggedright\begin{thebibliography}{100}

\bibitem{Misner:1969hg}
C.W.~Misner, \emph{{Mixmaster universe}},
  \href{https://doi.org/10.1103/PhysRevLett.22.1071}{\emph{Phys. Rev. Lett.}
  {\bfseries 22} (1969) 1071}.

\bibitem{Kasner:1921zz}
E.~Kasner, \emph{{Geometrical theorems on Einstein's cosmological equations}},
  \href{https://doi.org/10.2307/2370192}{\emph{Am. J. Math.} {\bfseries 43}
  (1921) 217}.

\bibitem{Belinsky:1970ew}
V.A.~Belinsky, I.M.~Khalatnikov and E.M.~Lifshitz, \emph{{Oscillatory approach
  to a singular point in the relativistic cosmology}},
  \href{https://doi.org/10.1080/00018737000101171}{\emph{Adv. Phys.} {\bfseries
  19} (1970) 525}.

\bibitem{Belinski:2017fas}
V.~Belinski and M.~Henneaux, \emph{The Cosmological Singularity}, Cambridge
  Monographs on Mathematical Physics, Cambridge University Press, Cambridge, UK
  (2017), \href{https://doi.org/10.1017/9781107239333}{10.1017/9781107239333}.

\bibitem{Belinski:2009wj}
V.~Belinski, \emph{{Cosmological singularity}},
  \href{https://doi.org/10.1063/1.3382327}{\emph{AIP Conf. Proc.} {\bfseries
  1205} (2010) 17} [\href{https://arxiv.org/abs/0910.0374}{{\ttfamily
  arXiv:0910.0374}}].

\bibitem{Belinski:2014kba}
V.A.~Belinski, \emph{{On the cosmological singularity}},
  \href{https://doi.org/10.1142/S021827181430016X}{\emph{Int. J. Mod. Phys. D}
  {\bfseries 23} (2014) 1430016}
  [\href{https://arxiv.org/abs/1404.3864}{{\ttfamily arXiv:1404.3864}}].

\bibitem{Damour:2000wm}
T.~Damour and M.~Henneaux, \emph{{Chaos in superstring cosmology}},
  \href{https://doi.org/10.1103/PhysRevLett.85.920}{\emph{Phys. Rev. Lett.}
  {\bfseries 85} (2000) 920}
  [\href{https://arxiv.org/abs/hep-th/0003139}{{\ttfamily hep-th/0003139}}].

\bibitem{Damour:2000hv}
T.~Damour and M.~Henneaux, \emph{{E(10), BE(10) and arithmetical chaos in
  superstring cosmology}},
  \href{https://doi.org/10.1103/PhysRevLett.86.4749}{\emph{Phys. Rev. Lett.}
  {\bfseries 86} (2001) 4749}
  [\href{https://arxiv.org/abs/hep-th/0012172}{{\ttfamily hep-th/0012172}}].

\bibitem{Damour:2002tc}
T.~Damour, M.~Henneaux, A.D.~Rendall and M.~Weaver, \emph{{Kasner like behavior
  for subcritical Einstein matter systems}},
  \href{https://doi.org/10.1007/s000230200000}{\emph{Annales Henri Poincar\'e}
  {\bfseries 3} (2002) 1049}
  [\href{https://arxiv.org/abs/gr-qc/0202069}{{\ttfamily gr-qc/0202069}}].

\bibitem{Damour:2002et}
T.~Damour, M.~Henneaux and H.~Nicolai, \emph{{Cosmological billiards}},
  \href{https://doi.org/10.1088/0264-9381/20/9/201}{\emph{Class. Quant. Grav.}
  {\bfseries 20} (2003) R145}
  [\href{https://arxiv.org/abs/hep-th/0212256}{{\ttfamily hep-th/0212256}}].

\bibitem{Middleton:2008rh}
J.~Middleton and J.D.~Barrow, \emph{{The Stability of an Isotropic Cosmological
  Singularity in Higher-Order Gravity}},
  \href{https://doi.org/10.1103/PhysRevD.77.103523}{\emph{Phys. Rev. D}
  {\bfseries 77} (2008) 103523}
  [\href{https://arxiv.org/abs/0801.4090}{{\ttfamily arXiv:0801.4090}}].

\bibitem{Barrow:2005qv}
J.D.~Barrow and S.~Hervik, \emph{{Anisotropically inflating universes}},
  \href{https://doi.org/10.1103/PhysRevD.73.023007}{\emph{Phys. Rev. D}
  {\bfseries 73} (2006) 023007}
  [\href{https://arxiv.org/abs/gr-qc/0511127}{{\ttfamily gr-qc/0511127}}].

\bibitem{Barrow:2006xb}
J.D.~Barrow and S.~Hervik, \emph{{On the evolution of universes in quadratic
  theories of gravity}},
  \href{https://doi.org/10.1103/PhysRevD.74.124017}{\emph{Phys. Rev. D}
  {\bfseries 74} (2006) 124017}
  [\href{https://arxiv.org/abs/gr-qc/0610013}{{\ttfamily gr-qc/0610013}}].

\bibitem{Sakakihara:2020rdy}
Y.~Sakakihara, D.~Yoshida, K.~Takahashi and J.~Quintin, \emph{{Theories with
  limited extrinsic curvature and a nonsingular anisotropic universe}},
  \href{https://doi.org/10.1103/PhysRevD.102.084004}{\emph{Phys. Rev. D}
  {\bfseries 102} (2020) 084004}
  [\href{https://arxiv.org/abs/2005.10844}{{\ttfamily arXiv:2005.10844}}].

\bibitem{Starobinsky:2019xdp}
A.A.~Starobinsky, S.V.~Sushkov and M.S.~Volkov, \emph{{Anisotropy screening in
  Horndeski cosmologies}},
  \href{https://doi.org/10.1103/PhysRevD.101.064039}{\emph{Phys. Rev. D}
  {\bfseries 101} (2020) 064039}
  [\href{https://arxiv.org/abs/1912.12320}{{\ttfamily arXiv:1912.12320}}].

\bibitem{Galeev:2021xit}
R.~Galeev, R.~Muharlyamov, A.A.~Starobinsky, S.V.~Sushkov and M.S.~Volkov,
  \emph{{Anisotropic cosmological models in Horndeski gravity}},
  \href{https://doi.org/10.1103/PhysRevD.103.104015}{\emph{Phys. Rev. D}
  {\bfseries 103} (2021) 104015}
  [\href{https://arxiv.org/abs/2102.10981}{{\ttfamily arXiv:2102.10981}}].

\bibitem{Stelle:1976gc}
K.S.~Stelle, \emph{{Renormalization of Higher Derivative Quantum Gravity}},
  \href{https://doi.org/10.1103/PhysRevD.16.953}{\emph{Phys. Rev. D} {\bfseries
  16} (1977) 953}.

\bibitem{Lehners:2019ibe}
J.L.~Lehners and K.S.~Stelle, \emph{{A Safe Beginning for the Universe?}},
  \href{https://doi.org/10.1103/PhysRevD.100.083540}{\emph{Phys. Rev. D}
  {\bfseries 100} (2019) 083540}
  [\href{https://arxiv.org/abs/1909.01169}{{\ttfamily arXiv:1909.01169}}].

\bibitem{Jonas:2021xkx}
C.~Jonas, J.L.~Lehners and J.~Quintin, \emph{{Cosmological consequences of a
  principle of finite amplitudes}},
  \href{https://doi.org/10.1103/PhysRevD.103.103525}{\emph{Phys. Rev. D}
  {\bfseries 103} (2021) 103525}
  [\href{https://arxiv.org/abs/2102.05550}{{\ttfamily arXiv:2102.05550}}].

\bibitem{Erickson:2003zm}
J.K.~Erickson, D.H.~Wesley, P.J.~Steinhardt and N.~Turok, \emph{{Kasner and
  mixmaster behavior in universes with equation of state $w \geq 1$}},
  \href{https://doi.org/10.1103/PhysRevD.69.063514}{\emph{Phys. Rev. D}
  {\bfseries 69} (2004) 063514}
  [\href{https://arxiv.org/abs/hep-th/0312009}{{\ttfamily hep-th/0312009}}].

\bibitem{Garfinkle:2008ei}
D.~Garfinkle, W.C.~Lim, F.~Pretorius and P.J.~Steinhardt, \emph{{Evolution to a
  smooth universe in an ekpyrotic contracting phase with $w > 1$}},
  \href{https://doi.org/10.1103/PhysRevD.78.083537}{\emph{Phys. Rev. D}
  {\bfseries 78} (2008) 083537}
  [\href{https://arxiv.org/abs/0808.0542}{{\ttfamily arXiv:0808.0542}}].

\bibitem{Cook:2020oaj}
W.G.~Cook, I.A.~Glushchenko, A.~Ijjas, F.~Pretorius and P.J.~Steinhardt,
  \emph{{Supersmoothing through Slow Contraction}},
  \href{https://doi.org/10.1016/j.physletb.2020.135690}{\emph{Phys. Lett. B}
  {\bfseries 808} (2020) 135690}
  [\href{https://arxiv.org/abs/2006.01172}{{\ttfamily arXiv:2006.01172}}].

\bibitem{Ijjas:2020dws}
A.~Ijjas, W.G.~Cook, F.~Pretorius, P.J.~Steinhardt and E.Y.~Davies,
  \emph{{Robustness of slow contraction to cosmic initial conditions}},
  \href{https://doi.org/10.1088/1475-7516/2020/08/030}{\emph{JCAP} {\bfseries
  08} (2020) 030} [\href{https://arxiv.org/abs/2006.04999}{{\ttfamily
  arXiv:2006.04999}}].

\bibitem{Ijjas:2021gkf}
A.~Ijjas, A.P.~Sullivan, F.~Pretorius, P.J.~Steinhardt and W.G.~Cook,
  \emph{{Ultralocality and Slow Contraction}},
  \href{https://doi.org/10.1088/1475-7516/2021/06/013}{\emph{JCAP} {\bfseries
  06} (2021) 013} [\href{https://arxiv.org/abs/2103.00584}{{\ttfamily
  arXiv:2103.00584}}].

\bibitem{Ijjas:2021wml}
A.~Ijjas, F.~Pretorius, P.J.~Steinhardt and A.P.~Sullivan, \emph{{The effects
  of multiple modes and reduced symmetry on the rapidity and robustness of slow
  contraction}},
  \href{https://doi.org/10.1016/j.physletb.2021.136490}{\emph{Phys. Lett. B}
  {\bfseries 820} (2021) 136490}
  [\href{https://arxiv.org/abs/2104.12293}{{\ttfamily arXiv:2104.12293}}].

\bibitem{Lidsey:2005wr}
J.E.~Lidsey, \emph{{Cosmic no hair for collapsing universes}},
  \href{https://doi.org/10.1088/0264-9381/23/10/018}{\emph{Class. Quant. Grav.}
  {\bfseries 23} (2006) 3517}
  [\href{https://arxiv.org/abs/hep-th/0511174}{{\ttfamily hep-th/0511174}}].

\bibitem{Barrow:2015wfa}
J.D.~Barrow and C.~Ganguly, \emph{{Evolution of initially contracting Bianchi
  Class A models in the presence of an ultra-stiff anisotropic pressure
  fluid}}, \href{https://doi.org/10.1088/0264-9381/33/12/125004}{\emph{Class.
  Quant. Grav.} {\bfseries 33} (2016) 125004}
  [\href{https://arxiv.org/abs/1510.01095}{{\ttfamily arXiv:1510.01095}}].

\bibitem{Wands:1998yp}
D.~Wands, \emph{{Duality invariance of cosmological perturbation spectra}},
  \href{https://doi.org/10.1103/PhysRevD.60.023507}{\emph{Phys. Rev. D}
  {\bfseries 60} (1999) 023507}
  [\href{https://arxiv.org/abs/gr-qc/9809062}{{\ttfamily gr-qc/9809062}}].

\bibitem{Finelli:2001sr}
F.~Finelli and R.~Brandenberger, \emph{{On the generation of a scale invariant
  spectrum of adiabatic fluctuations in cosmological models with a contracting
  phase}}, \href{https://doi.org/10.1103/PhysRevD.65.103522}{\emph{Phys. Rev.
  D} {\bfseries 65} (2002) 103522}
  [\href{https://arxiv.org/abs/hep-th/0112249}{{\ttfamily hep-th/0112249}}].

\bibitem{Brandenberger:2012zb}
R.H.~Brandenberger, \emph{{The Matter Bounce Alternative to Inflationary
  Cosmology}},  \href{https://arxiv.org/abs/1206.4196}{{\ttfamily
  arXiv:1206.4196}}.

\bibitem{Levy:2016xcl}
A.M.~Levy, \emph{{Fine-tuning challenges for the matter bounce scenario}},
  \href{https://doi.org/10.1103/PhysRevD.95.023522}{\emph{Phys. Rev. D}
  {\bfseries 95} (2017) 023522}
  [\href{https://arxiv.org/abs/1611.08972}{{\ttfamily arXiv:1611.08972}}].

\bibitem{Lin:2017fec}
C.~Lin, J.~Quintin and R.H.~Brandenberger, \emph{{Massive gravity and the
  suppression of anisotropies and gravitational waves in a matter-dominated
  contracting universe}},
  \href{https://doi.org/10.1088/1475-7516/2018/01/011}{\emph{JCAP} {\bfseries
  01} (2018) 011} [\href{https://arxiv.org/abs/1711.10472}{{\ttfamily
  arXiv:1711.10472}}].

\bibitem{Bramberger:2019zez}
S.F.~Bramberger and J.L.~Lehners, \emph{{Nonsingular bounces catalyzed by dark
  energy}}, \href{https://doi.org/10.1103/PhysRevD.99.123523}{\emph{Phys. Rev.
  D} {\bfseries 99} (2019) 123523}
  [\href{https://arxiv.org/abs/1901.10198}{{\ttfamily arXiv:1901.10198}}].

\bibitem{Anabalon:2019equ}
A.~Anabal\'on, S.F.~Bramberger and J.L.~Lehners, \emph{{Kerr-NUT-de Sitter as
  an Inhomogeneous Non-Singular Bouncing Cosmology}},
  \href{https://doi.org/10.1007/JHEP09(2019)096}{\emph{JHEP} {\bfseries 09}
  (2019) 096} [\href{https://arxiv.org/abs/1904.07285}{{\ttfamily
  arXiv:1904.07285}}].

\bibitem{Kumar:2021mgc}
K.S.~Kumar, S.~Maheshwari, A.~Mazumdar and J.~Peng, \emph{{An anisotropic
  bouncing universe in non-local gravity}},
  \href{https://doi.org/10.1088/1475-7516/2021/07/025}{\emph{JCAP} {\bfseries
  07} (2021) 025} [\href{https://arxiv.org/abs/2103.13980}{{\ttfamily
  arXiv:2103.13980}}].

\bibitem{Rajeev:2021yyl}
K.~Rajeev, V.~Mondal and S.~Chakraborty, \emph{{Bouncing with shear:
  Implications from quantum cosmology}},
  \href{https://arxiv.org/abs/2109.08696}{{\ttfamily arXiv:2109.08696}}.

\bibitem{Planck:2018jri}
{\scshape Planck} collaboration, Y.~Akrami et~al., \emph{{Planck 2018 results.
  X. Constraints on inflation}},
  \href{https://doi.org/10.1051/0004-6361/201833887}{\emph{Astron. Astrophys.}
  {\bfseries 641} (2020) A10}
  [\href{https://arxiv.org/abs/1807.06211}{{\ttfamily arXiv:1807.06211}}].

\bibitem{Faraoni:2021lfc}
V.~Faraoni and A.~Giusti, \emph{{Thermodynamics of scalar-tensor gravity}},
  \href{https://doi.org/10.1103/PhysRevD.103.L121501}{\emph{Phys. Rev. D}
  {\bfseries 103} (2021) L121501}
  [\href{https://arxiv.org/abs/2103.05389}{{\ttfamily arXiv:2103.05389}}].

\bibitem{Giusti:2021sku}
A.~Giusti, S.~Zentarra, L.~Heisenberg and V.~Faraoni, \emph{{First-order
  thermodynamics of Horndeski gravity}},
  \href{https://arxiv.org/abs/2108.10706}{{\ttfamily arXiv:2108.10706}}.

\bibitem{Misner:1967zz}
C.W.~Misner, \emph{{Neutrino Viscosity and the Isotropy of Primordial Blackbody
  Radiation}}, \href{https://doi.org/10.1103/PhysRevLett.19.533}{\emph{Phys.
  Rev. Lett.} {\bfseries 19} (1967) 533}.

\bibitem{Misner:1967uu}
C.W.~Misner, \emph{{The Isotropy of the universe}},
  \href{https://doi.org/10.1086/149448}{\emph{Astrophys. J.} {\bfseries 151}
  (1968) 431}.

\bibitem{Stewart:1968}
J.M.~{Stewart}, \emph{{Neutrino Viscosity in Cosmological Models}},
  {\emph{Astrophysical Letters} {\bfseries 2} (1968) 133}.

\bibitem{Matzner:1969}
R.A.~{Matzner}, \emph{{The Evolution of Anisotropy in Nonrotating Bianchi Type
  V Cosmologies}}, \href{https://doi.org/10.1086/150138}{\emph{Astrophys. J.}
  {\bfseries 157} (1969) 1085}.

\bibitem{Weinberg:1971mx}
S.~Weinberg, \emph{{Entropy generation and the survival of protogalaxies in an
  expanding universe}}, \href{https://doi.org/10.1086/151073}{\emph{Astrophys.
  J.} {\bfseries 168} (1971) 175}.

\bibitem{Matzner:1972b}
R.A.~{Matzner}, \emph{{Dissipative Effects in the Expansion of the Universe.
  II. a Multicomponent Model for Neutrino Dissipation of Anisotropy in the
  Early Universe}}, \href{https://doi.org/10.1086/151295}{\emph{Astrophys. J.}
  {\bfseries 171} (1972) 433}.

\bibitem{Weinberg:2003ur}
S.~Weinberg, \emph{{Damping of tensor modes in cosmology}},
  \href{https://doi.org/10.1103/PhysRevD.69.023503}{\emph{Phys. Rev. D}
  {\bfseries 69} (2004) 023503}
  [\href{https://arxiv.org/abs/astro-ph/0306304}{{\ttfamily
  astro-ph/0306304}}].

\bibitem{Hawking:1966qi}
S.W.~Hawking, \emph{{Perturbations of an expanding universe}},
  \href{https://doi.org/10.1086/148793}{\emph{Astrophys. J.} {\bfseries 145}
  (1966) 544}.

\bibitem{Stewart:1969}
J.M.~{Stewart}, \emph{{Non-equilibrium processes in the early universe}},
  \href{https://doi.org/10.1093/mnras/145.3.347}{\emph{Mon. Not. Roy. Astron.
  Soc.} {\bfseries 145} (1969) 347}.

\bibitem{Matzner:1972}
R.A.~{Matzner} and C.W.~{Misner}, \emph{{Dissipative Effects in the Expansion
  of the Universe. I.}}, \href{https://doi.org/10.1086/151294}{\emph{Astrophys.
  J.} {\bfseries 171} (1972) 415}.

\bibitem{Parnovskii:1977}
S.L.~{Parnovski{\v{i}}}, \emph{{Influence of viscosity on the evolution of a
  Bianchi type II universe}}, {\emph{JETP} {\bfseries 45} (1977) 423}.

\bibitem{Belinskii:1979}
V.A.~{Belinski{\v{i}}}, E.S.~{Nikomarov} and I.M.~{Khalatnikov},
  \emph{{Investigation of the cosmological evolution of viscoelastic matter
  with causal thermodynamics}}, {\emph{JETP} {\bfseries 50} (1979) 213}.

\bibitem{Gron:1990ew}
{\O}.~Gr\o{}n, \emph{{Viscous inflationary universe models}},
  \href{https://doi.org/10.1007/BF00643930}{\emph{Astrophys. Space Sci.}
  {\bfseries 173} (1990) 191}.

\bibitem{Hervik}
{\O}.~Gr\o{}n and S.~Hervik, \emph{Einstein's General Theory of Relativity},
  Springer-Verlag New York (2007),
  \href{https://doi.org/10.1007/978-0-387-69200-5}{10.1007/978-0-387-69200-5}.

\bibitem{Brevik:review}
I.~Brevik, {\O}.~Gr\o{}n, J.~de~Haro, S.D.~Odintsov and E.N.~Saridakis,
  \emph{{Viscous Cosmology for Early- and Late-Time Universe}},
  \href{https://doi.org/10.1142/S0218271817300245}{\emph{Int. J. Mod. Phys. D}
  {\bfseries 26} (2017) 1730024}
  [\href{https://arxiv.org/abs/1706.02543}{{\ttfamily arXiv:1706.02543}}].

\bibitem{Brevik:2019yma}
I.~Brevik and S.~Nojiri, \emph{{Gravitational Waves in the Presence of
  Viscosity}}, \href{https://doi.org/10.1142/S0218271819501335}{\emph{Int. J.
  Mod. Phys. D} {\bfseries 28} (2019) 1950133}
  [\href{https://arxiv.org/abs/1901.00767}{{\ttfamily arXiv:1901.00767}}].

\bibitem{Anand:2017wsj}
S.~Anand, P.~Chaubal, A.~Mazumdar and S.~Mohanty, \emph{{Cosmic viscosity as a
  remedy for tension between PLANCK and LSS data}},
  \href{https://doi.org/10.1088/1475-7516/2017/11/005}{\emph{JCAP} {\bfseries
  11} (2017) 005} [\href{https://arxiv.org/abs/1708.07030}{{\ttfamily
  arXiv:1708.07030}}].

\bibitem{Goswami:2016tsu}
G.~Goswami, G.K.~Chakravarty, S.~Mohanty and A.R.~Prasanna, \emph{{Constraints
  on cosmological viscosity and self interacting dark matter from gravitational
  wave observations}},
  \href{https://doi.org/10.1103/PhysRevD.95.103509}{\emph{Phys. Rev. D}
  {\bfseries 95} (2017) 103509}
  [\href{https://arxiv.org/abs/1603.02635}{{\ttfamily arXiv:1603.02635}}].

\bibitem{Lu:2018smr}
B.Q.~Lu, D.~Huang, Y.L.~Wu and Y.F.~Zhou, \emph{{Damping of gravitational waves
  in a viscous Universe and its implication for dark matter
  self-interactions}},  \href{https://arxiv.org/abs/1803.11397}{{\ttfamily
  arXiv:1803.11397}}.

\bibitem{Atreya:2017pny}
A.~Atreya, J.R.~Bhatt and A.~Mishra, \emph{{Viscous Self Interacting Dark
  Matter and Cosmic Acceleration}},
  \href{https://doi.org/10.1088/1475-7516/2018/02/024}{\emph{JCAP} {\bfseries
  02} (2018) 024} [\href{https://arxiv.org/abs/1709.02163}{{\ttfamily
  arXiv:1709.02163}}].

\bibitem{Natwariya:2019fif}
P.K.~Natwariya, J.R.~Bhatt and A.K.~Pandey, \emph{{Viscosity in cosmic
  fluids}}, \href{https://doi.org/10.1140/epjc/s10052-020-8341-8}{\emph{Eur.
  Phys. J. C} {\bfseries 80} (2020) 767}
  [\href{https://arxiv.org/abs/1907.03445}{{\ttfamily arXiv:1907.03445}}].

\bibitem{Mishra:2020onx}
A.K.~Mishra, \emph{{Exploring the Self Interacting Dark Matter Properties From
  Low Redshift Observations}},
  \href{https://arxiv.org/abs/2002.11652}{{\ttfamily arXiv:2002.11652}}.

\bibitem{Belinski:2013jua}
V.~Belinski, \emph{{Stabilization of the Friedmann big bang by the shear
  stresses}}, \href{https://doi.org/10.1103/PhysRevD.88.103521}{\emph{Phys.
  Rev. D} {\bfseries 88} (2013) 103521}
  [\href{https://arxiv.org/abs/1310.5112}{{\ttfamily arXiv:1310.5112}}].

\bibitem{Ganguly:2019llh}
C.~Ganguly and M.~Bruni, \emph{{Quasi-Isotropic Cycles and Nonsingular Bounces
  in a Mixmaster Cosmology}},
  \href{https://doi.org/10.1103/PhysRevLett.123.201301}{\emph{Phys. Rev. Lett.}
  {\bfseries 123} (2019) 201301}
  [\href{https://arxiv.org/abs/1902.06356}{{\ttfamily arXiv:1902.06356}}].

\bibitem{Ganguly:2020daq}
C.~Ganguly, \emph{Cosmic no-hair theorems for viscous contracting universes},
  \href{https://doi.org/10.1088/1475-7516/2021/07/013}{\emph{JCAP} {\bfseries
  2021} (2021) 013} [\href{https://arxiv.org/abs/2008.02286}{{\ttfamily
  arXiv:2008.02286}}].

\bibitem{Eckart:1940te}
C.~Eckart, \emph{{The Thermodynamics of irreversible processes. 3..
  Relativistic theory of the simple fluid}},
  \href{https://doi.org/10.1103/PhysRev.58.919}{\emph{Phys. Rev.} {\bfseries
  58} (1940) 919}.

\bibitem{LandauLifshitz}
L.~Landau and E.~Lifshitz, \emph{Fluid Mechanics}, second~ed., Pergamon,
  Oxford, UK (1987),
  \href{https://doi.org/10.1016/C2013-0-03799-1}{10.1016/C2013-0-03799-1}.

\bibitem{Israel:1979wp}
W.~Israel and J.M.~Stewart, \emph{{Transient relativistic thermodynamics and
  kinetic theory}},
  \href{https://doi.org/10.1016/0003-4916(79)90130-1}{\emph{Annals Phys.}
  {\bfseries 118} (1979) 341}.

\bibitem{Jeon:1994if}
S.~Jeon, \emph{{Hydrodynamic transport coefficients in relativistic scalar
  field theory}}, \href{https://doi.org/10.1103/PhysRevD.52.3591}{\emph{Phys.
  Rev. D} {\bfseries 52} (1995) 3591}
  [\href{https://arxiv.org/abs/hep-ph/9409250}{{\ttfamily hep-ph/9409250}}].

\bibitem{Jeon:1995zm}
S.~Jeon and L.G.~Yaffe, \emph{{From quantum field theory to hydrodynamics:
  Transport coefficients and effective kinetic theory}},
  \href{https://doi.org/10.1103/PhysRevD.53.5799}{\emph{Phys. Rev. D}
  {\bfseries 53} (1996) 5799}
  [\href{https://arxiv.org/abs/hep-ph/9512263}{{\ttfamily hep-ph/9512263}}].

\bibitem{Kapusta:2006pm}
J.~Kapusta and C.~Gale, \emph{{Finite-temperature field theory: Principles and
  applications}}, Cambridge Monographs on Mathematical Physics, Cambridge
  University Press, Cambridge, UK (2011),
  \href{https://doi.org/10.1017/CBO9780511535130}{10.1017/CBO9780511535130}.

\bibitem{Policastro:2001yc}
G.~Policastro, D.T.~Son and A.O.~Starinets, \emph{{The Shear viscosity of
  strongly coupled N=4 supersymmetric Yang-Mills plasma}},
  \href{https://doi.org/10.1103/PhysRevLett.87.081601}{\emph{Phys. Rev. Lett.}
  {\bfseries 87} (2001) 081601}
  [\href{https://arxiv.org/abs/hep-th/0104066}{{\ttfamily hep-th/0104066}}].

\bibitem{Kovtun:2004de}
P.~Kovtun, D.T.~Son and A.O.~Starinets, \emph{{Viscosity in strongly
  interacting quantum field theories from black hole physics}},
  \href{https://doi.org/10.1103/PhysRevLett.94.111601}{\emph{Phys. Rev. Lett.}
  {\bfseries 94} (2005) 111601}
  [\href{https://arxiv.org/abs/hep-th/0405231}{{\ttfamily hep-th/0405231}}].

\bibitem{Son:2007vk}
D.T.~Son and A.O.~Starinets, \emph{{Viscosity, Black Holes, and Quantum Field
  Theory}},
  \href{https://doi.org/10.1146/annurev.nucl.57.090506.123120}{\emph{Ann. Rev.
  Nucl. Part. Sci.} {\bfseries 57} (2007) 95}
  [\href{https://arxiv.org/abs/0704.0240}{{\ttfamily arXiv:0704.0240}}].

\bibitem{Banks:2002fe}
T.~Banks and W.~Fischler, \emph{{Black crunch}},
  \href{https://arxiv.org/abs/hep-th/0212113}{{\ttfamily hep-th/0212113}}.

\bibitem{Quintin:2016qro}
J.~Quintin and R.H.~Brandenberger, \emph{{Black hole formation in a contracting
  universe}}, \href{https://doi.org/10.1088/1475-7516/2016/11/029}{\emph{JCAP}
  {\bfseries 11} (2016) 029}
  [\href{https://arxiv.org/abs/1609.02556}{{\ttfamily arXiv:1609.02556}}].

\bibitem{Chen:2016kjx}
J.W.~Chen, J.~Liu, H.L.~Xu and Y.F.~Cai, \emph{{Tracing Primordial Black Holes
  in Nonsingular Bouncing Cosmology}},
  \href{https://doi.org/10.1016/j.physletb.2017.03.036}{\emph{Phys. Lett. B}
  {\bfseries 769} (2017) 561}
  [\href{https://arxiv.org/abs/1609.02571}{{\ttfamily arXiv:1609.02571}}].

\bibitem{Lehners:2008qe}
J.L.~Lehners and P.J.~Steinhardt, \emph{{Dark Energy and the Return of the
  Phoenix Universe}},
  \href{https://doi.org/10.1103/PhysRevD.79.063503}{\emph{Phys. Rev. D}
  {\bfseries 79} (2009) 063503}
  [\href{https://arxiv.org/abs/0812.3388}{{\ttfamily arXiv:0812.3388}}].

\bibitem{Lehners:2009eg}
J.L.~Lehners, P.J.~Steinhardt and N.~Turok, \emph{{The Return of the Phoenix
  Universe}}, \href{https://doi.org/10.1142/S0218271809015977}{\emph{Int. J.
  Mod. Phys. D} {\bfseries 18} (2009) 2231}
  [\href{https://arxiv.org/abs/0910.0834}{{\ttfamily arXiv:0910.0834}}].

\bibitem{Masoumi:2014vpa}
A.~Masoumi and S.D.~Mathur, \emph{{An equation of state in the limit of high
  densities}}, \href{https://doi.org/10.1103/PhysRevD.90.084052}{\emph{Phys.
  Rev. D} {\bfseries 90} (2014) 084052}
  [\href{https://arxiv.org/abs/1406.5798}{{\ttfamily arXiv:1406.5798}}].

\bibitem{Masoumi:2015sga}
A.~Masoumi, \emph{{State of matter at high density and entropy bounds}},
  \href{https://doi.org/10.1142/S0218271815440162}{\emph{Int. J. Mod. Phys. D}
  {\bfseries 24} (2015) 1544016}
  [\href{https://arxiv.org/abs/1505.06787}{{\ttfamily arXiv:1505.06787}}].

\bibitem{Masoumi:2014nfa}
A.~Masoumi and S.D.~Mathur, \emph{{A violation of the covariant entropy
  bound?}}, \href{https://doi.org/10.1103/PhysRevD.91.084058}{\emph{Phys. Rev.
  D} {\bfseries 91} (2015) 084058}
  [\href{https://arxiv.org/abs/1412.2618}{{\ttfamily arXiv:1412.2618}}].

\bibitem{Mathur:2020ivc}
S.D.~Mathur, \emph{{Three puzzles in cosmology}},
  \href{https://doi.org/10.1142/S021827182030013X}{\emph{Int. J. Mod. Phys. D}
  {\bfseries 29} (2020) 2030013}
  [\href{https://arxiv.org/abs/2009.09832}{{\ttfamily arXiv:2009.09832}}].

\bibitem{Banks:2001px}
T.~Banks and W.~Fischler, \emph{{An Holographic cosmology}},
  \href{https://arxiv.org/abs/hep-th/0111142}{{\ttfamily hep-th/0111142}}.

\bibitem{Banks:2003ta}
T.~Banks and W.~Fischler, \emph{{Holographic cosmology 3.0}},
  \href{https://doi.org/10.1238/Physica.Topical.117a00056}{\emph{Phys. Scripta
  T} {\bfseries 117} (2005) 56}
  [\href{https://arxiv.org/abs/hep-th/0310288}{{\ttfamily hep-th/0310288}}].

\bibitem{Banks:2004cw}
T.~Banks, W.~Fischler and L.~Mannelli, \emph{{Microscopic quantum mechanics of
  the p = rho universe}},
  \href{https://doi.org/10.1103/PhysRevD.71.123514}{\emph{Phys. Rev. D}
  {\bfseries 71} (2005) 123514}
  [\href{https://arxiv.org/abs/hep-th/0408076}{{\ttfamily hep-th/0408076}}].

\bibitem{Banks:2004vg}
T.~Banks and W.~Fischler, \emph{{Holographic cosmology}},
  \href{https://arxiv.org/abs/hep-th/0405200}{{\ttfamily hep-th/0405200}}.

\bibitem{Veneziano:2003sz}
G.~Veneziano, \emph{{A Model for the big bounce}},
  \href{https://doi.org/10.1088/1475-7516/2004/03/004}{\emph{JCAP} {\bfseries
  03} (2004) 004} [\href{https://arxiv.org/abs/hep-th/0312182}{{\ttfamily
  hep-th/0312182}}].

\bibitem{Quintin:2018loc}
J.~Quintin, R.H.~Brandenberger, M.~Gasperini and G.~Veneziano, \emph{{Stringy
  black-hole gas in \ensuremath{\alpha}'-corrected dilaton gravity}},
  \href{https://doi.org/10.1103/PhysRevD.98.103519}{\emph{Phys. Rev. D}
  {\bfseries 98} (2018) 103519}
  [\href{https://arxiv.org/abs/1809.01658}{{\ttfamily arXiv:1809.01658}}].

\bibitem{Carr:2011hv}
B.J.~Carr and A.A.~Coley, \emph{{Persistence of black holes through a
  cosmological bounce}},
  \href{https://doi.org/10.1142/S0218271811020640}{\emph{Int. J. Mod. Phys. D}
  {\bfseries 20} (2011) 2733}
  [\href{https://arxiv.org/abs/1104.3796}{{\ttfamily arXiv:1104.3796}}].

\bibitem{Clifton:2017hvg}
T.~Clifton, B.~Carr and A.~Coley, \emph{{Persistent Black Holes in Bouncing
  Cosmologies}}, \href{https://doi.org/10.1088/1361-6382/aa6dbb}{\emph{Class.
  Quant. Grav.} {\bfseries 34} (2017) 135005}
  [\href{https://arxiv.org/abs/1701.05750}{{\ttfamily arXiv:1701.05750}}].

\bibitem{Carr:2017wkz}
B.~Carr, T.~Clifton and A.~Coley, \emph{{Black holes as echoes of previous
  cosmic cycles}},  \href{https://arxiv.org/abs/1704.02919}{{\ttfamily
  arXiv:1704.02919}}.

\bibitem{Coley:2020ykx}
A.A.~Coley, \emph{{Persistence in black hole lattice cosmological models}},
  \href{https://doi.org/10.1088/1361-6382/abbf31}{\emph{Class. Quant. Grav.}
  {\bfseries 37} (2020) 245002}
  [\href{https://arxiv.org/abs/2012.14049}{{\ttfamily arXiv:2012.14049}}].

\bibitem{Rovelli:2018hbk}
C.~Rovelli and F.~Vidotto, \emph{{White-hole dark matter and the origin of past
  low-entropy}},  \href{https://arxiv.org/abs/1804.04147}{{\ttfamily
  arXiv:1804.04147}}.

\bibitem{Rovelli:2018hba}
C.~Rovelli and F.~Vidotto, \emph{{Pre-Big-Bang Black-Hole Remnants and Past Low
  Entropy}}, \href{https://doi.org/10.3390/universe4110129}{\emph{Universe}
  {\bfseries 4} (2018) 129} [\href{https://arxiv.org/abs/1805.03224}{{\ttfamily
  arXiv:1805.03224}}].

\bibitem{Barrau:2021spy}
A.~Barrau, L.~Ferdinand, K.~Martineau and C.~Renevey, \emph{{Closer look at
  white hole remnants}},
  \href{https://doi.org/10.1103/PhysRevD.103.043532}{\emph{Phys. Rev. D}
  {\bfseries 103} (2021) 043532}
  [\href{https://arxiv.org/abs/2101.01949}{{\ttfamily arXiv:2101.01949}}].

\bibitem{LeTiec:2020spy}
A.~Le~Tiec and M.~Casals, \emph{{Spinning Black Holes Fall in Love}},
  \href{https://doi.org/10.1103/PhysRevLett.126.131102}{\emph{Phys. Rev. Lett.}
  {\bfseries 126} (2021) 131102}
  [\href{https://arxiv.org/abs/2007.00214}{{\ttfamily arXiv:2007.00214}}].

\bibitem{Chia:2020yla}
H.S.~Chia, \emph{{Tidal deformation and dissipation of rotating black holes}},
  \href{https://doi.org/10.1103/PhysRevD.104.024013}{\emph{Phys. Rev. D}
  {\bfseries 104} (2021) 024013}
  [\href{https://arxiv.org/abs/2010.07300}{{\ttfamily arXiv:2010.07300}}].

\bibitem{Charalambous:2021mea}
P.~Charalambous, S.~Dubovsky and M.M.~Ivanov, \emph{{On the Vanishing of Love
  Numbers for Kerr Black Holes}},
  \href{https://doi.org/10.1007/JHEP05(2021)038}{\emph{JHEP} {\bfseries 05}
  (2021) 038} [\href{https://arxiv.org/abs/2102.08917}{{\ttfamily
  arXiv:2102.08917}}].

\bibitem{Ehlers:1993gf}
J.~Ehlers, \emph{{Contributions to the relativistic mechanics of continuous
  media}}, \href{https://doi.org/10.1007/BF00759031}{\emph{Gen. Rel. Grav.}
  {\bfseries 25} (1993) 1225}.

\bibitem{ellis_maartens_maccallum_2012}
G.F.R.~Ellis, R.~Maartens and M.A.H.~MacCallum, \emph{Relativistic Cosmology},
  Cambridge University Press, Cambridge, UK (2012),
  \href{https://doi.org/10.1017/CBO9781139014403}{10.1017/CBO9781139014403}.

\bibitem{Israel:1976tn}
W.~Israel, \emph{{Nonstationary irreversible thermodynamics: A Causal
  relativistic theory}},
  \href{https://doi.org/10.1016/0003-4916(76)90064-6}{\emph{Annals Phys.}
  {\bfseries 100} (1976) 310}.

\bibitem{LandauLifshitz2}
L.~Landau and E.~Lifshitz, \emph{Theory of Elasticity}, second~ed., Pergamon,
  Oxford, UK (1970),
  \href{https://doi.org/10.1016/C2009-0-25521-8}{10.1016/C2009-0-25521-8}.

\bibitem{Bozza:2009jx}
V.~Bozza and M.~Bruni, \emph{{A Solution to the anisotropy problem in bouncing
  cosmologies}},
  \href{https://doi.org/10.1088/1475-7516/2009/10/014}{\emph{JCAP} {\bfseries
  10} (2009) 014} [\href{https://arxiv.org/abs/0909.5611}{{\ttfamily
  arXiv:0909.5611}}].

\bibitem{Ananda:2005xp}
K.N.~Ananda and M.~Bruni, \emph{{Cosmo-dynamics and dark energy with non-linear
  equation of state: a quadratic model}},
  \href{https://doi.org/10.1103/PhysRevD.74.023523}{\emph{Phys. Rev. D}
  {\bfseries 74} (2006) 023523}
  [\href{https://arxiv.org/abs/astro-ph/0512224}{{\ttfamily
  astro-ph/0512224}}].

\bibitem{Ananda:2006gf}
K.N.~Ananda and M.~Bruni, \emph{{Cosmo-dynamics and dark energy with a
  quadratic EoS: Anisotropic models, large-scale perturbations and cosmological
  singularities}},
  \href{https://doi.org/10.1103/PhysRevD.74.023524}{\emph{Phys. Rev. D}
  {\bfseries 74} (2006) 023524}
  [\href{https://arxiv.org/abs/gr-qc/0603131}{{\ttfamily gr-qc/0603131}}].

\bibitem{ChapmanCowling}
S.~{Chapman} and T.G.~{Cowling}, \emph{{The mathematical theory of non-uniform
  gases: an account of the kinetic theory of viscosity, thermal conduction and
  diffusion in gases}}, Cambridge Mathematical Library, Cambridge University
  Press, Cambridge, UK (1991).

\bibitem{LeBellac}
M.~Le~Bellac, F.~Mortessagne and G.G.~Batrouni, \emph{Equilibrium and
  Non-Equilibrium Statistical Thermodynamics}, Cambridge University Press,
  Cambridge, UK (2004),
  \href{https://doi.org/10.1017/CBO9780511606571}{10.1017/CBO9780511606571}.

\bibitem{Burshtein}
A.I.~{Burshtein}, \emph{{Introduction to Thermodynamics and Kinetic Theory of
  Matter}}, 2nd~ed., Wiley-VCH, Weinheim, Germany (2005).

\bibitem{1972JETP...34.1159Z}
Y.B.~{Zel'Dovich} and A.A.~{Starobinski{\v{i}}}, \emph{{Particle Production and
  Vacuum Polarization in an Anisotropic Gravitational Field}}, {\emph{JETP}
  {\bfseries 34} (1972) 1159}.

\bibitem{Allahverdi:2005mz}
R.~Allahverdi and A.~Mazumdar, \emph{{Supersymmetric thermalization and
  quasi-thermal universe: Consequences for gravitinos and leptogenesis}},
  \href{https://doi.org/10.1088/1475-7516/2006/10/008}{\emph{JCAP} {\bfseries
  10} (2006) 008} [\href{https://arxiv.org/abs/hep-ph/0512227}{{\ttfamily
  hep-ph/0512227}}].

\bibitem{Loeb:2020lwa}
A.~Loeb, \emph{{Upper Limit on the Dissipation of Gravitational Waves in
  Gravitationally Bound Systems}},
  \href{https://doi.org/10.3847/2041-8213/ab72ab}{\emph{Astrophys. J. Lett.}
  {\bfseries 890} (2020) L16}
  [\href{https://arxiv.org/abs/2001.01730}{{\ttfamily arXiv:2001.01730}}].

\bibitem{Hawking:1973uf}
S.W.~Hawking and G.F.R.~Ellis, \emph{{The Large Scale Structure of
  Space-Time}}, Cambridge Monographs on Mathematical Physics, Cambridge
  University Press (2011),
  \href{https://doi.org/10.1017/CBO9780511524646}{10.1017/CBO9780511524646}.

\bibitem{Kiefer:2018uyv}
C.~Kiefer, N.~Kwidzinski and W.~Piechocki, \emph{{On the dynamics of the
  general Bianchi IX spacetime near the singularity}},
  \href{https://doi.org/10.1140/epjc/s10052-018-6155-8}{\emph{Eur. Phys. J. C}
  {\bfseries 78} (2018) 691}
  [\href{https://arxiv.org/abs/1807.06261}{{\ttfamily arXiv:1807.06261}}].

\bibitem{lukashGW}
V.N.~{Lukash}, \emph{{Homogeneous cosmological models with gravitational waves
  and rotation}}, {\emph{JETP Letters} {\bfseries 19} (1974) 265}.

\bibitem{Pritchard:2004qp}
J.R.~Pritchard and M.~Kamionkowski, \emph{{Cosmic microwave background
  fluctuations from gravitational waves: An Analytic approach}},
  \href{https://doi.org/10.1016/j.aop.2005.03.005}{\emph{Annals Phys.}
  {\bfseries 318} (2005) 2}
  [\href{https://arxiv.org/abs/astro-ph/0412581}{{\ttfamily
  astro-ph/0412581}}].

\bibitem{Watanabe:2006qe}
Y.~Watanabe and E.~Komatsu, \emph{{Improved Calculation of the Primordial
  Gravitational Wave Spectrum in the Standard Model}},
  \href{https://doi.org/10.1103/PhysRevD.73.123515}{\emph{Phys. Rev. D}
  {\bfseries 73} (2006) 123515}
  [\href{https://arxiv.org/abs/astro-ph/0604176}{{\ttfamily
  astro-ph/0604176}}].

\bibitem{Stefanek:2012hj}
B.A.~Stefanek and W.W.~Repko, \emph{{Analytic description of the damping of
  gravitational waves by free streaming neutrinos}},
  \href{https://doi.org/10.1103/PhysRevD.88.083536}{\emph{Phys. Rev. D}
  {\bfseries 88} (2013) 083536}
  [\href{https://arxiv.org/abs/1207.7285}{{\ttfamily arXiv:1207.7285}}].

\bibitem{Dent:2013asa}
J.B.~Dent, L.M.~Krauss, S.~Sabharwal and T.~Vachaspati, \emph{{Damping of
  Primordial Gravitational Waves from Generalized Sources}},
  \href{https://doi.org/10.1103/PhysRevD.88.084008}{\emph{Phys. Rev. D}
  {\bfseries 88} (2013) 084008}
  [\href{https://arxiv.org/abs/1307.7571}{{\ttfamily arXiv:1307.7571}}].

\bibitem{Baym:2017xvh}
G.~Baym, S.P.~Patil and C.J.~Pethick, \emph{{Damping of gravitational waves by
  matter}}, \href{https://doi.org/10.1103/PhysRevD.96.084033}{\emph{Phys. Rev.
  D} {\bfseries 96} (2017) 084033}
  [\href{https://arxiv.org/abs/1707.05192}{{\ttfamily arXiv:1707.05192}}].

\bibitem{Kite:2021yoe}
T.~Kite, J.~Chluba, A.~Ravenni and S.P.~Patil, \emph{{Clarifying transfer
  function approximations for the large-scale gravitational wave background in
  $\Lambda$CDM}},  \href{https://arxiv.org/abs/2107.13351}{{\ttfamily
  arXiv:2107.13351}}.

\bibitem{1992ApJ...395...34B}
M.~{Bruni}, P.K.S.~{Dunsby} and G.F.R.~{Ellis}, \emph{{Cosmological
  Perturbations and the Physical Meaning of Gauge-invariant Variables}},
  \href{https://doi.org/10.1086/171629}{\emph{\apj} {\bfseries 395} (1992) 34}.

\bibitem{Pereira:2019mpp}
T.S.~Pereira and C.~Pitrou, \emph{{Bianchi spacetimes as supercurvature modes
  around isotropic cosmologies}},
  \href{https://doi.org/10.1103/PhysRevD.100.123534}{\emph{Phys. Rev. D}
  {\bfseries 100} (2019) 123534}
  [\href{https://arxiv.org/abs/1909.13688}{{\ttfamily arXiv:1909.13688}}].

\bibitem{Fanizza:2021ngq}
G.~Fanizza, M.~Gasperini, E.~Pavone and L.~Tedesco, \emph{{Linearized
  propagation equations for metric fluctuations in a general (non-vacuum)
  background geometry}},
  \href{https://doi.org/10.1088/1475-7516/2021/07/021}{\emph{JCAP} {\bfseries
  07} (2021) 021} [\href{https://arxiv.org/abs/2105.13041}{{\ttfamily
  arXiv:2105.13041}}].

\bibitem{Quintin:2015rta}
J.~Quintin, Z.~Sherkatghanad, Y.F.~Cai and R.H.~Brandenberger, \emph{{Evolution
  of cosmological perturbations and the production of non-Gaussianities through
  a nonsingular bounce: Indications for a no-go theorem in single field matter
  bounce cosmologies}},
  \href{https://doi.org/10.1103/PhysRevD.92.063532}{\emph{Phys. Rev. D}
  {\bfseries 92} (2015) 063532}
  [\href{https://arxiv.org/abs/1508.04141}{{\ttfamily arXiv:1508.04141}}].

\bibitem{Li:2016xjb}
Y.B.~Li, J.~Quintin, D.G.~Wang and Y.F.~Cai, \emph{{Matter bounce cosmology
  with a generalized single field: non-Gaussianity and an extended no-go
  theorem}}, \href{https://doi.org/10.1088/1475-7516/2017/03/031}{\emph{JCAP}
  {\bfseries 03} (2017) 031}
  [\href{https://arxiv.org/abs/1612.02036}{{\ttfamily arXiv:1612.02036}}].

\bibitem{1974JETP...39..742L}
V.N.~{Lukash} and A.A.~{Starobinski{\v{i}}}, \emph{{The isotropization of the
  cosmological expansion owing to particle production}}, {\emph{JETP}
  {\bfseries 39} (1974) 742}.

\bibitem{Hu:1978zd}
B.L.~Hu and L.~Parker, \emph{{Anisotropy Damping Through Quantum Effects in the
  Early Universe}}, \href{https://doi.org/10.1103/PhysRevD.17.933}{\emph{Phys.
  Rev. D} {\bfseries 17} (1978) 933}.

\bibitem{Hartle:1980nn}
J.B.~Hartle and B.L.~Hu, \emph{{Quantum effects in the early universe. III.
  Dissipation of anisotropy by scalar particle production}},
  \href{https://doi.org/10.1103/PhysRevD.21.2756}{\emph{Phys. Rev. D}
  {\bfseries 21} (1980) 2756}.

\bibitem{Calzetta:1986ey}
E.~Calzetta and B.L.~Hu, \emph{{Closed Time Path Functional Formalism in Curved
  Space-Time: Application to Cosmological Back Reaction Problems}},
  \href{https://doi.org/10.1103/PhysRevD.35.495}{\emph{Phys. Rev. D} {\bfseries
  35} (1987) 495}.

\bibitem{Mottola:1985ee}
E.~Mottola, \emph{{A Quantum Fluctuation Dissipation Theorem for General
  Relativity}}, \href{https://doi.org/10.1103/PhysRevD.33.2136}{\emph{Phys.
  Rev. D} {\bfseries 33} (1986) 2136}.

\bibitem{Cai:2011zx}
Y.F.~Cai, R.~Brandenberger and X.~Zhang, \emph{{The Matter Bounce Curvaton
  Scenario}}, \href{https://doi.org/10.1088/1475-7516/2011/03/003}{\emph{JCAP}
  {\bfseries 03} (2011) 003} [\href{https://arxiv.org/abs/1101.0822}{{\ttfamily
  arXiv:1101.0822}}].

\bibitem{Qiu:2015nha}
T.~Qiu and Y.T.~Wang, \emph{{G-Bounce Inflation: Towards Nonsingular Inflation
  Cosmology with Galileon Field}},
  \href{https://doi.org/10.1007/JHEP04(2015)130}{\emph{JHEP} {\bfseries 04}
  (2015) 130} [\href{https://arxiv.org/abs/1501.03568}{{\ttfamily
  arXiv:1501.03568}}].

\bibitem{Graham:2019bfu}
P.W.~Graham, D.E.~Kaplan and S.~Rajendran, \emph{{Relaxation of the
  Cosmological Constant}},
  \href{https://doi.org/10.1103/PhysRevD.100.015048}{\emph{Phys. Rev. D}
  {\bfseries 100} (2019) 015048}
  [\href{https://arxiv.org/abs/1902.06793}{{\ttfamily arXiv:1902.06793}}].

\bibitem{Ji:2021mvg}
L.~Ji, D.E.~Kaplan, S.~Rajendran and E.H.~Tanin, \emph{{Thermal Perturbations
  from Cosmological Constant Relaxation}},
  \href{https://arxiv.org/abs/2109.05285}{{\ttfamily arXiv:2109.05285}}.

\bibitem{Nandi:2019xag}
D.~Nandi and L.~Sriramkumar, \emph{{Can a nonminimal coupling restore the
  consistency condition in bouncing universes?}},
  \href{https://doi.org/10.1103/PhysRevD.101.043506}{\emph{Phys. Rev. D}
  {\bfseries 101} (2020) 043506}
  [\href{https://arxiv.org/abs/1904.13254}{{\ttfamily arXiv:1904.13254}}].

\bibitem{Nandi:2020sif}
D.~Nandi, \emph{{Bounce from Inflation}},
  \href{https://doi.org/10.1016/j.physletb.2020.135695}{\emph{Phys. Lett. B}
  {\bfseries 809} (2020) 135695}
  [\href{https://arxiv.org/abs/2003.02066}{{\ttfamily arXiv:2003.02066}}].

\bibitem{Nandi:2020szp}
D.~Nandi, \emph{{Stability of a viable non-minimal bounce}},
  \href{https://doi.org/10.3390/universe7030062}{\emph{Universe} {\bfseries 7}
  (2021) 62} [\href{https://arxiv.org/abs/2009.03134}{{\ttfamily
  arXiv:2009.03134}}].

\bibitem{Akama:2019qeh}
S.~Akama, S.~Hirano and T.~Kobayashi, \emph{{Primordial non-Gaussianities of
  scalar and tensor perturbations in general bounce cosmology: Evading the
  no-go theorem}},
  \href{https://doi.org/10.1103/PhysRevD.101.043529}{\emph{Phys. Rev. D}
  {\bfseries 101} (2020) 043529}
  [\href{https://arxiv.org/abs/1908.10663}{{\ttfamily arXiv:1908.10663}}].

\bibitem{Pereira:2007yy}
T.S.~Pereira, C.~Pitrou and J.P.~Uzan, \emph{{Theory of cosmological
  perturbations in an anisotropic universe}},
  \href{https://doi.org/10.1088/1475-7516/2007/09/006}{\emph{JCAP} {\bfseries
  09} (2007) 006} [\href{https://arxiv.org/abs/0707.0736}{{\ttfamily
  arXiv:0707.0736}}].

\bibitem{Ed}
I.~Agullo, J.~Olmedo and E.~Wilson-Ewing, to appear.

\bibitem{Grain:2020wro}
J.~Grain and V.~Vennin, \emph{{Unavoidable shear from quantum fluctuations in
  contracting cosmologies}},
  \href{https://doi.org/10.1140/epjc/s10052-021-08932-0}{\emph{Eur. Phys. J. C}
  {\bfseries 81} (2021) 132}
  [\href{https://arxiv.org/abs/2005.04222}{{\ttfamily arXiv:2005.04222}}].

\bibitem{deJong:2021bbo}
E.~de~Jong, J.C.~Aurrekoetxea and E.A.~Lim, \emph{{Primordial black hole
  formation with full numerical relativity}},
  \href{https://arxiv.org/abs/2109.04896}{{\ttfamily arXiv:2109.04896}}.

\bibitem{Bousso:1997wi}
R.~Bousso and S.W.~Hawking, \emph{{(Anti)evaporation of Schwarzschild-de Sitter
  black holes}}, \href{https://doi.org/10.1103/PhysRevD.57.2436}{\emph{Phys.
  Rev. D} {\bfseries 57} (1998) 2436}
  [\href{https://arxiv.org/abs/hep-th/9709224}{{\ttfamily hep-th/9709224}}].

\bibitem{Gregory:2018ghc}
R.~Gregory, D.~Kastor and J.~Traschen, \emph{{Evolving Black Holes in
  Inflation}}, \href{https://doi.org/10.1088/1361-6382/aacec2}{\emph{Class.
  Quant. Grav.} {\bfseries 35} (2018) 155008}
  [\href{https://arxiv.org/abs/1804.03462}{{\ttfamily arXiv:1804.03462}}].

\bibitem{Kaloper:2010ec}
N.~Kaloper, M.~Kleban and D.~Martin, \emph{{McVittie's Legacy: Black Holes in
  an Expanding Universe}},
  \href{https://doi.org/10.1103/PhysRevD.81.104044}{\emph{Phys. Rev. D}
  {\bfseries 81} (2010) 104044}
  [\href{https://arxiv.org/abs/1003.4777}{{\ttfamily arXiv:1003.4777}}].

\bibitem{Faraoni:2012gz}
V.~Faraoni, A.F.~Zambrano~Moreno and R.~Nandra, \emph{{Making sense of the
  bizarre behaviour of horizons in the McVittie spacetime}},
  \href{https://doi.org/10.1103/PhysRevD.85.083526}{\emph{Phys. Rev. D}
  {\bfseries 85} (2012) 083526}
  [\href{https://arxiv.org/abs/1202.0719}{{\ttfamily arXiv:1202.0719}}].

\bibitem{Faraoni:2013aba}
V.~Faraoni, \emph{{Evolving black hole horizons in General Relativity and
  alternative gravity}},
  \href{https://doi.org/10.3390/galaxies1030114}{\emph{Galaxies} {\bfseries 1}
  (2013) 114} [\href{https://arxiv.org/abs/1309.4915}{{\ttfamily
  arXiv:1309.4915}}].

\bibitem{Barrau:2017ukm}
A.~Barrau, K.~Martineau and F.~Moulin, \emph{{Seeing through the cosmological
  bounce: Footprints of the contracting phase and luminosity distance in
  bouncing models}},
  \href{https://doi.org/10.1103/PhysRevD.96.123520}{\emph{Phys. Rev. D}
  {\bfseries 96} (2017) 123520}
  [\href{https://arxiv.org/abs/1711.05301}{{\ttfamily arXiv:1711.05301}}].

\bibitem{Papanikolaou:2020qtd}
T.~Papanikolaou, V.~Vennin and D.~Langlois, \emph{{Gravitational waves from a
  universe filled with primordial black holes}},
  \href{https://doi.org/10.1088/1475-7516/2021/03/053}{\emph{JCAP} {\bfseries
  03} (2021) 053} [\href{https://arxiv.org/abs/2010.11573}{{\ttfamily
  arXiv:2010.11573}}].

\end{thebibliography}\endgroup

\end{document}